\documentclass[useAMS,usenatbib]{mn2e}
\usepackage{graphicx,amsmath}
\usepackage{amssymb}
\usepackage{natbib}
\usepackage{color}
\usepackage[colorlinks=true,
            citecolor=blue, 
            linkcolor = blue,
            urlcolor  = blue,
            citecolor = blue,
            anchorcolor = blue]{hyperref}
\usepackage{hyperref}
\title[Redshift evolution of escape fraction ]{The redshift evolution of escape fraction of hydrogen ionizing photons from galaxies }
\author[Khaire et al]
{
\parbox{\textwidth}{ 
Vikram Khaire$^{1}$ \thanks{E-mail:vikramk@iucaa.in},  
Raghunathan Srianand$^{1}$,
Tirthankar Roy Choudhury$^{2}$,\\
and Prakash Gaikwad$^{2}$ 
} 
\vspace*{10pt}\\ 
$^{1}$Inter-University Centre for Astronomy and Astrophysics (IUCAA), Post Bag 4, Pune 411007, India\\
$^{2}$National Centre for Radio Astrophysics, Tata Institute of Fundamental Research, Pune 411007, India\\  
}   

\begin{document}

\defcitealias{Khaire15ebl}{KS15b}
\defcitealias{Khaire15puc}{KS15a}
\defcitealias{HM12}{HM12}
\defcitealias{Madau15}{MH15}

\date{Accepted 2016 January 20.}

\pagerange{\pageref{firstpage}--\pageref{lastpage}} \pubyear{2016}
\maketitle

\label{firstpage}

\vspace{20 mm}

\begin{abstract} 
Using our cosmological radiative transfer code, we study the implications 
of the updated quasi-stellar object (QSO) emissivity and star formation history for the escape 
fraction ($f_{\rm esc}$) of hydrogen ionizing photons from galaxies. 
We estimate the $f_{\rm esc}$ that is required to reionize the Universe and to 
maintain the ionization state of the intergalactic medium in 
the post-reionization era. At $z>5.5$, we show that a 
constant $f_{\rm esc}$ of 0.14 to 0.22 is sufficient to reionize the 
Universe. At $z<3.5$, consistent with various observations, we find that 
$f_{\rm esc}$ can have values from 0 to 0.05. However, a steep rise in 
$f_{\rm esc}$, of at least a factor of $\sim 3$, is required between 
$z=3.5$ to $5.5$. It results from a rapidly decreasing QSO emissivity 
at $z>3$ together with a nearly constant measured H~{\sc i} photoionization rates 
at $3<z<5$. We show that this requirement of a steep rise in $f_{\rm esc}$ 
over a very short time can be relaxed  if we consider the contribution 
from a recently found large number density of faint QSOs at $z\ge4$. 
In addition, a simple extrapolation of the contribution of such QSOs
to high-$z$ suggests that QSOs alone can reionize the Universe. 
This implies, at $z>3.5$, that either the properties of galaxies 
should evolve rapidly to increase the $f_{\rm esc}$ or most of the low-mass 
galaxies should host massive black holes and sustain accretion over a 
prolonged period. 
These results motivate a careful investigation of theoretical predictions
of these alternate scenarios
that can be distinguished using future observations. 
Moreover, it is also very important
to revisit the measurements of H~{\sc i} photoionization rates that are crucial 
to the analysis presented here. 
\end{abstract}


\section{Introduction}
The hydrogen in the intergalactic medium (IGM) is highly ionized at $z<6$ 
as shown by various observations of Lyman-$\alpha$ forest in high-redshift 
quasi-stellar object (QSO) spectra 
\citep{Fan06, Bolton11, Goto11} and the cosmic microwave background (CMB) 
polarization measurements \citep{Larson11}. Prior to that the IGM went 
through a phase transition from a completely neutral to a highly ionized 
state through the process of H~{\sc i} reionization.
In the most typical scenario, QSOs and the H~{\sc i} ionizing photons 
generated by the star-forming galaxies reionize the Universe and maintain the 
observed high ionization state of the IGM even after the epoch of 
reionization. However, the relative contribution of galaxies to the total 
budget of H~{\sc i} ionizing photons is highly uncertain. This is because the 
fraction ($f_{\rm esc}$) of H~{\sc i} ionizing photons generated by 
stellar population inside the galaxies that escapes out into the IGM is 
ill-constrained.

Over the past decade, various observational studies have reported 
the average $f_{\rm esc}$ ranging from 0.02 to 0.5 at $z\sim3$
\citep{Steidel01, Shapley06, Iwata09, Boutsia11, Nestor13, Micheva15}. The 
recent deep observations of galaxies have demonstrated that most of the 
previously reported detections of the escaping H~{\sc i} ionizing photons are 
spurious due to contaminations from the low-$z$ intervening galaxies 
\citep{Mostardi15, Siana15}. The average $f_{\rm esc}$ inferred from a sample 
of galaxies are dominated by a handful of galaxies with high $f_{\rm esc}$. 
At low-$z$, apart from few individual galaxies with high $f_{\rm esc}$ 
\citep[see for e.g. ][]{Borthakur14}, various observational studies found 
that a very small fraction of the H~{\sc i} ionizing photons 
do escape from 
galaxies and the derived upper limits on the average $f_{\rm esc}$ hardly 
exceed 0.05 \citep{Cowie09,Grimes09, Bridge10, Barger13, Leitet13}.

On the other hand, theoretical studies have suggested a wide range of 
$f_{\rm esc}$ from 0.01 to 1 \citep{Dove94, Ricotti00, Gnedin08, Kimm14, Roy15}.
There are various factors that influence the escape of H~{\sc i} ionizing 
photons such as the galaxy mass, morphology, supernova rates, composition of
interstellar medium and the gas distribution 
\citep{Fernandez11, Benson13, Kim13, Cen15}. In a standard picture where 
population of bright QSOs decline rapidly at $z>3$, one requires the contribution 
from galaxies to dominate the H~{\sc i} ionizing photon budget and drive the
H~{\sc i} reionization. Observations related to H~{\sc i} reionization, such as 
galaxy luminosity functions at high redshift \citep{Bouwens12, Finkelstein15}, 
hydrogen photoionization rates inferred from the Lyman-$\alpha$ forest seen in 
the QSO absorption spectra, and constraints on the electron scattering optical 
depth ($\tau_{\rm el}$) from CMB 
\citep[e.g. from Wilkinson Microwave Anisotropy Probe; WMAP][]{Hinshaw13} have been 
used for constraining the $f_{\rm esc}$ at $z>6$. An evolving 
$f_{\rm esc}$ at high-z was necessary to match the WMAP 
$\tau_{\rm el}=0.089\pm 0.014$ \citep{HM12, Kuhlen12, Shull12, Mitra13}
however, no such evolution is required to explain the recent 
$\tau_{\rm el} = 0.066\pm0.016$ from \citet{Planck15}. Recently, \citet{Mitra15} 
have carried out a Markov chain Monte Carlo based statistical analysis using a 
semi-analytical model of reionization to infer that the current data sets can 
be explained by a constant $f_{\rm esc}\sim 0.15$ at $z > 6$, though the 
uncertainties on the constraints are still high. \citet{Bouwens15reion} also find 
similar constraints on $f_{\rm esc}$ by computing the evolution of the ionizing
emissivity from the galaxy luminosity functions.

The $f_{\rm esc}$ controls the ionizing emissivity of galaxies which is crucial 
to understand the H~{\sc i} reionization, thermal history of the IGM, and the 
ratio of column densities of He~{\sc ii} to H~{\sc i} inferred in the IGM 
\citep{Khaire13}. The $f_{\rm esc}$ is also important to study the implications 
of the trapped radiation in galaxies to understand the detectability of 
Lyman-$\alpha$ emission from primordial galaxies \citep{Tumlinson01, Rhoads04}.

In this paper, we study the effect of updated QSO emissivity and star formation 
rate density (SFRD) on the required mean $f_{\rm esc}$ to keep the IGM 
ionized at the level required by different observations.
We use the QSO emissivity from \citet{Khaire15puc} (hereafter, 
\citetalias{Khaire15puc}) which is obtained from a 
compilation of recent QSO
luminosity functions (QLFs). We use a self-consistently calculated 
SFRD along with the dust attenuation from \citet{Khaire15ebl} (hereafter, 
\citetalias{Khaire15ebl}) obtained by compiling various multi-wavelength and 
multi-epoch galaxy luminosity functions. To constrain $f_{\rm esc}$ in the 
post-reionization era ($z \lesssim 6$), we use the measured H~{\sc i} photoionization 
rates \citep{Bolton07, Calverley11, Wyithe11, Becker13, Kollmeier14, Shull15} and 
in the pre-reionization era ($z\gtrsim 6$), we calculate reionization histories 
consistent with the $\tau_{\rm el}$ constraints from Planck along with the recent 
mean H~{\sc i} fraction measurements \citep{Schenker14, McGreer15}. We calculate 
the H~{\sc i} ionizing UV background (UVB; in this paper it implies a background 
radiation at $\lambda<912$~\AA) using a radiative transfer code developed by us 
\citep{Khaire13} following the standard prescription 
\citep{Miralda90, Shapiro94, HM96, Fardal98, Shull99, FG09} albeit keeping
$f_{\rm esc}$ as a free parameter. Then, we constrain $f_{\rm esc}(z)$ by comparing 
the model predictions with the observations mentioned above. We also study the implications of high QSO 
emissivity at $z>4$ obtained using the recent QLFs reported by \citet{Giallongo15} 
on the $f_{\rm esc}$ and H~{\sc i} reionization. 

The plan of the paper is as follows. In Section~\ref{sec2}, we present the basic 
theory to evaluate the UVB in pre- and post-reionization era. We review the form 
of H~{\sc i} ionizing emissivity from QSOs and galaxies used in our models in 
Section~\ref{sec3}. In Section~\ref{sec4}, we summarize the constraints on 
$f_{\rm esc}$ in the post- and pre-reionization era for different possible QSO 
emissivities as a function of $z$. We also explore an extreme case where all the 
H~{\sc i} ionizing photons required for the reionization are sourced by QSOs 
alone (i.e with $f_{\rm esc}=0$). In Section~\ref{sec5}, we summarize the results.
Throughout this paper we use a flat $\rm \Lambda$CDM cosmology with $\Omega_{\Lambda}=0.7$, 
$\Omega_{m}=0.3$ and $H_0=70$ km s$^{-1}$ Mpc$^{-1}$.
\section{Basic theory}\label{sec2}
The definition of escape fraction ($f_{\rm esc}$) of H~{\sc i} ionizing photons 
from galaxies used in this paper is
%
\begin{equation}\label{Eq.def}
f_{\rm esc}=\frac{L(912{\rm\AA})_{\rm esc}}{L(912{\rm\AA})_{\rm int}}\,\,,
\end{equation} 
%
where, $L(912{\rm\AA})_{\rm esc}$ and $L(912{\rm\AA})_{\rm int}$ are the 
escaping and intrinsic specific luminosities at $\lambda=912$~\AA, respectively.
The procedure we use to estimate the  mean $f_{\rm esc}$ from galaxies at all 
epochs is briefly described here. We estimate the mean specific intensity 
$J_{\nu}$(z) of the H~{\sc i} ionizing UVB (contributed by both QSOs and 
galaxies) by taking $f_{\rm esc}(z)$ as a free parameter. Using this, we 
obtain H~{\sc i} photoionization rates ($\Gamma_{\rm HI}$), defined as 
%
\begin{equation}\label{Eq.gama}
\Gamma_{\rm HI} (z) = \int^{\infty}_{\nu_{912}}{\frac{4\pi J_{\nu}(z)\, 
\sigma_{\rm HI}(\nu) d\nu}{h \nu}}\,\,,
\end{equation} 
%
where $\nu_{912}$ is the frequency of photons having energy equal to the 
ionization potential of H~{\sc i} (13.6 eV or $912$~\AA), $\sigma_{\rm HI}(\nu)$ is the 
H~{\sc i} photoionization cross-section and $h$ is the Planck constant. To 
constrain $f_{\rm esc}(z)$, in the post-reionization era, we compare predicted 
$\Gamma_{\rm HI}(z)$ from our models with its measurements \citep[see][for a 
similar method]{Inoue06} and in the pre-reionization era, we compare our model 
predictions with the recent constraints on $\tau_{\rm el}$ from Planck and the 
mean fraction of H~{\sc i}.

The method described above depends crucially on the the H~{\sc i} ionizing UVB.
The UVB at any point in the Universe is the mean intensity of UV radiation 
arriving at that point generated by the relevant sources and filtered through 
gas in the IGM. Therefore, apart from the ionizing emissivity of the radiating 
sources, the distribution of gas in the IGM is also essential for computing the 
UVB. In this section, we provide the basic theory required to estimate the UVB 
and describe the models used for the IGM gas distribution in both pre- and 
post-reionization era. 
\subsection{UV background at all epochs}\label{sec1}
In the post-reionization era, the mean free path of H~{\sc i} ionizing photons 
is large enough to consider the UVB as essentially uniform. In the pre-reionization 
era, each source or the cluster of sources of the H~{\sc i} ionizing photons create 
a bubble of H~{\sc ii} around them. The volume filling factor ($Q_{\rm HII}$) of 
all these  H~{\sc ii} bubbles, defined as the ratio of the volume occupied by all 
such H~{\sc ii} bubbles to the total volume of the Universe, is crucial for 
understanding the dynamics of the H~{\sc i} reionization. The $Q_{\rm HII}$, is 
usually obtained under the assumption that the interior of all such H~{\sc ii} 
bubbles have, on an average, a uniform UVB 
\citep[however, see][for local fluctuations in the UVB]{Meiksin03}. Therefore, in 
the pre-reionization era, the UVB is within the H~{\sc ii} bubbles and there are 
no H~{\sc i} ionizing UV photons outside the bubbles \citep[see][]{Choudhury09}. 
The $Q_{\rm HII}(z)$ is equal to unity in the post-reionization era and the exact 
value of it ($Q_{\rm HII}(z)<1$) in the pre-reionization era depends on the model 
of the H~{\sc i} reionization. Taking all these into consideration, the mean 
specific intensity of UVB, $J_{\nu_0}$ (in units of 
erg cm$^{-2}$ s$^{-1}$ Hz$^{-1}$ sr$^{-1}$), at all epochs $z_0$ and frequency 
$\nu_0$ can be evaluated by solving the following integral 
\citep[see][]{Peebles93, HM96},
%
\begin{equation}\label{Eq.uvb}
J_{\nu_{0}}(z_{0})=\frac{c}{4\pi}\int_{z_{0}}^{\infty}\,dz\frac{(1+z_{0})^{3}\,
\epsilon_{\nu}(z)}{(1+z)H(z)Q_{\rm HII}(z)} e^{-\tau_{\rm eff}(\nu_{0},z_{0},z)}.
\end{equation}
%
Here, the $H(z)=H_0 \sqrt{\Omega_m(1+z)^3+\Omega_{\Lambda}}$ is the Hubble 
parameter, $c$ is the speed of light and $\epsilon_{\nu}(z)$ is the comoving 
H~{\sc i} ionizing emissivity of the sources. 
The $Q_{\rm HII}(z)$ in the denominator makes our UVB different
from the UVB model of \citet[][hereafter, \citetalias{HM12}]{HM12} in the pre-reionization 
era. This
ensures that the  H~{\sc i} ionizing photons are confined within the H~{\sc ii} 
bubbles in the pre-reionization era.\footnote{Note that, 
the $Q_{\rm HII}(z)$ in the denominator can give rise to 
very high values of $J_{\nu}(z)$ and $\Gamma_{\rm HI}(z)$ when 
it is very small.} This approximation of uniform UVB within the 
H~{\sc ii} bubbles breaks down at very small values of $Q_{\rm HII}$.
The $\tau_{\rm eff}(\nu_{0},z_{0},z)$ is an 
effective optical depth encountered by the ionizing photons while travelling from 
the emission redshift $z\,>z_0$ with a frequency $\nu\,>\nu_0$ to arrive at redshift 
$z_0$ and frequency $\nu_0$. Therefore, the relation between $\nu$ and $\nu_0$ is 
given by $\nu=\nu_0(1+z)/(1+z_0)$.

For a Poisson distributed H~{\sc i} clouds in the IGM,  the $\tau_{\rm eff}$ is 
given by \citep[see][]{Paresce, Paddy3},
%
\begin{equation}\label{Eq.taueff}
\tau_{\rm eff}(\nu_{0}, z_{0}, z)=\int_{z_{0}}^{z}dz'\int_{0}^{\infty}dN_{\rm HI}
f(N_{\rm HI}, z') (1-e^{-\tau_{\nu'}})\,.
\end{equation}
%
Here, $f(N_{\rm HI}, z')=\frac{d^2N}{dN_{\rm HI}dz'}$ is the number of H~{\sc i} 
clouds per unit redshift and column density interval $N_{\rm HI}$ to  
$N_{\rm HI} + dN_{\rm HI} $. The continuum optical depth $\tau_{\nu'}$ is 
%
\begin{equation}\label{Eq.tauc}
\tau_{\nu'}=N_{\rm HI}{\sigma_{\rm HI}(\nu')}+N_{\rm HeI}{\sigma_{\rm HeI}(\nu')} 
+N_{\rm HeII}{\sigma_{\rm HeII}(\nu')},
\end{equation}
%
where, $\nu'=\nu_{0}(1+z')/(1+z_{0})$, and $N_i$ and $\sigma_{i}$ are the column 
density and photoionization cross-section, respectively, for species $i$. 
The ionization potential of He~{\sc i} (24.6 eV) is close to that of H~{\sc i}, and
total helium is an order of magnitude less abundant by number 
compare to hydrogen. Therefore,
the contribution of He~{\sc i} to the continuum optical depth is negligible. We do 
not consider this term (second term in Eq.~\ref{Eq.tauc}) further in our analysis.

In the post-reionization era, to calculate $\tau_{\rm eff}$ 
we use the $f(N_{\rm HI}, z)$  given by
\citet{InoueAK14}. It is consistent with the various measurements of
the H~{\sc i} column density distributions from Lyman-$\alpha$ forest \citep{Kim13}, 
Lyman limit systems \citep{Prochaska14, OMeara13, OMeara07} and damped 
Lyman-$\alpha$ systems \citep{Noterdaeme09, Noterdaeme12},  the mean free path 
of H~{\sc i} ionizing photons \citep{Worseck14LLS} and the opacity of the IGM to 
Lyman-$\alpha$ photons \citep{Fan06, Becker13Ly}. To calculate the UVB at wavelength 
$\lambda <228$~\AA, we follow the prescription given in \citetalias{HM12}
to obtain the ratio of $N_{\rm He II}$ to $N_{\rm H I}$. 

In the pre-reionization era, the helium within the H~{\sc ii} bubbles created by 
galaxies is mostly in He~{\sc ii}. Consequently, the UVB at $\lambda <228$~\AA~is 
negligible owing to its photo-absorption by He~{\sc ii}. This may not be true for 
the H~{\sc ii} bubbles created by bright QSOs. However, this wavelength range 
($\lambda <228$~\AA) has negligible effect on the $\Gamma_{\rm HI}$ because of 
the $\nu^{-3}$ dependence of the $\sigma_{\rm HI}(\nu)$. Therefore, in the 
pre-reionization era, we simplify the Eq.~(\ref{Eq.tauc}) to 
$\tau_{\nu'}=N_{\rm HI}{\sigma_{\rm HI}(\nu')}$ and estimate the UVB at 
$\lambda >228$~\AA. However, we still need the $f(N_{\rm HI}, z)$ that can not be 
directly measured due to large Gunn-Peterson optical depths at high-$z$. We obtain 
$f(N_{\rm HI}, z)$ by using a large number of random lines of sight generated through 
a high-resolution hydrodynamical simulation box as explained in the following section. 
\subsection{$f(N_{\rm HI}, z)$ within H~{\sc ii} bubble }\label{sec2.2}
We use a high-resolution hydrodynamic {\sc p-gadget3} simulation\footnote{The 
simulation uses $\sigma_8=0.827$, $\rm n_s=0.96$ and $\Omega_b=0.048$.} 
\citep{Springel05} with 2$\times$512$^3$ particles and a box size of 10 $h^{-1}$ 
cMpc to study the hydrogen distribution and to obtain the $f(N_{\rm HI}, z)$.
These boxes are generated from $z=9$ to $5$ with redshift interval of 0.5. In 
each box a large number of random lines of sight (LOS) were drawn. We use these 
LOS to probe the hydrogen distribution at different $z$. 

For a given $\Gamma_{\rm HI}$ in the simulation box, one can calculate the 
H~{\sc i} and H~{\sc ii} fraction assuming the photoionization equilibrium. 
However, to model the optically thin and thick absorption systems together, 
we need to consider the values of the local rate, 
$\Gamma^{\rm local}_{\rm HI}$, in regions with a range of over-densities. 
We use an empirical fit to the radiative transfer prescription of 
\citet{Rahmati13}, with a slight modification by \citet{Choudhury15}:
%
\begin{equation}\label{Eq.rahmati}
\frac{\Gamma^{\rm local}_{\rm HI}}{\Gamma_{\rm HI}}=
0.98\Bigg[1+\Big(\frac{\Delta_{\rm H}}{\Delta_{\rm ss}}\Big)^{1.64}\Bigg]^{-2.28} +
0.02\Bigg[1+\frac{\Delta_{\rm H}}{\Delta_{\rm ss}}\Bigg]^{-0.84}\,\,,
\end{equation}
%
where, $\Delta_{\rm H}$ is the overdensity of hydrogen and $\Delta_{\rm ss}$ 
is the threshold density. $\Delta_{\rm ss}$ is calculated assuming that the 
size of each absorber is given by the Jeans length \citep{Schaye01},
%
\begin{equation}\label{Eq.ss}
\begin{aligned}
\Delta_{\rm ss}=36 \Bigg( \frac{\Gamma_{\rm HI}}{10^{-12} 
{\rm s^{-1}}} \Bigg)^{2/3} \Bigg( \frac{T}{10^4 {\rm K}} \Bigg)^{2/15} \\
\Bigg( \frac{\mu}{0.61} \Bigg)^{1/3}\Bigg( \frac{\chi}{1.08} 
\Bigg)^{-2/3} \Bigg( \frac{1+z}{8} \Bigg)^{-3}\,\,,
\end{aligned}
\end{equation}
%
where, $T$ is the gas temperature, $\chi$ is the number of electrons per H~{\sc ii} 
ion in the H~{\sc ii} bubble and $\mu$ is the mean molecular weight. We take a 
constant $T=20000$K for the gas temperature 
within each H~{\sc ii} bubble. The value of $\Delta_{\rm ss}$ 
calculated in this way becomes unrealistically small for low $\Gamma_{\rm HI}$ at 
high $z$. Therefore, we impose a lower cut-off on $\Delta_{\rm ss}\ge 2$ 
\citep[see][]{Bolton13}. 

Using this prescription, we calculate the H~{\sc i} column density of the absorbers 
along each line of sight by integrating the $n_{\rm  HI}$ over a characteristic Jeans 
length ($L_{\rm ss}$) which is given by \citep{Bolton13}
%
\begin{equation}\label{Eq.nh1}
\begin{aligned}
L_{\rm ss}=8.7 {\rm pkpc} \Bigg( \frac{\Gamma_{\rm HI}}{10^{-12} 
{\rm s^{-1}}} \Bigg)^{-1/3} \Bigg( \frac{T}{10^4 {\rm K}} \Bigg)^{13/30} \\
\Bigg( \frac{\mu}{0.61} \Bigg)^{-2/3}\Bigg( \frac{\chi}{1.08} \Bigg)^{1/3}\,.
\end{aligned}
\end{equation}
%
\begin{figure}
\centering
\includegraphics[totalheight=0.31\textheight, trim=0cm 0cm 3.5cm 0cm, clip=true]{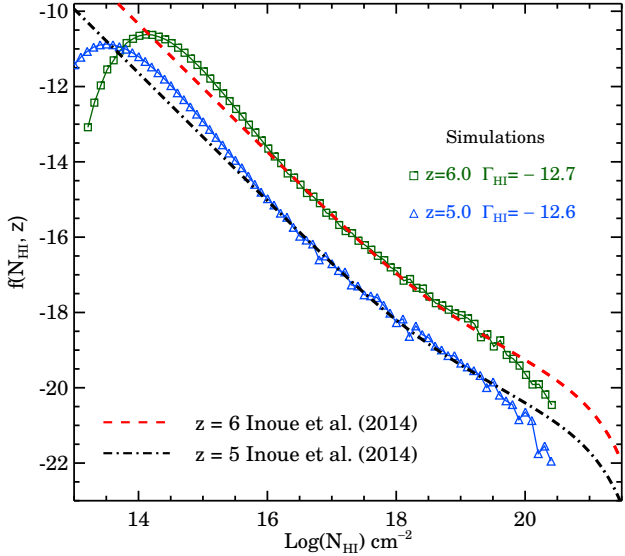}
\caption{The $f(N_{\rm HI}, z)$ at $z=5$ (\emph{blue triangles}) and $z=6$ (\emph{green squares}) 
obtained from the simulations using the method outlined in Section~\ref{sec2.2} 
are shown along with the fits obtained by \citet{InoueAK14} 
(\emph{black and red curves}) using various observations. The  $f(N_{\rm HI}, z)$ at 
$z=6$ is scaled up by factor 10 for the clarity in presentation. Over the relevant 
column density range for evaluating $\tau_{\rm eff}$ i.e, 
$19.5 >{\rm log} N_{\rm HI}> 15.5$, the $f(N_{\rm HI},z)$ matches quite well with 
the observations for the values of $\Gamma_{\rm HI}$ indicated in the plot. These 
$\Gamma_{\rm HI}$ values are chosen to be consistent with the measurements of 
\citet{Bolton07}, \citet{Calverley11} and \citet{Wyithe11}.}
\label{fig.fnh1z}
\end{figure}
%
It can have values between 20 to 40 pkpc in the redshift range of our interest 
($z>5.5$). We follow \citet{Bolton13} and use $L_{\rm ss}=20$ pkpc throughout to 
estimate the $f(N_{\rm HI}, z)$. For illustration, in Fig.~\ref{fig.fnh1z}, we show 
the $f(N_{\rm HI}, z)$ obtained in this way at two redshifts $z=5$ and $6$ for 
$\log (\Gamma_{\rm HI }\,{\rm s^{-1} })=-12.6$ and $-12.7$, respectively. These 
$\Gamma_{\rm HI}$ values are consistent with the measurements at corresponding $z$ 
\citep{Bolton07, Calverley11, Wyithe11}. We also show the empirical fits to the 
$f(N_{\rm HI}, z)$ given by \citet{InoueAK14}. At these redshifts, they match quite 
well with our estimates for $19.5 >{\rm log}(N_{\rm HI})> 15.5$. The $\tau_{\rm eff}$ 
is mostly dominated by this $N_{\rm HI}$ range. In the pre-reionization era, 
we use the $f(N_{\rm HI}, z)$ obtained from the simulations as 
explained above to estimate the UVB within the H~{\sc ii} bubbles.

In the following section, we discuss the basic theory to obtain the $Q_{\rm HII}(z)$ 
in the pre-reionization era.

\subsection{H~{\sc ii} volume filling factor }\label{sec1}
Under the assumption of the photoionization equilibrium within a H~{\sc ii} bubble, 
the time evolution of $Q_{\rm HII}$ can be written as \citep{Madau99, Barkana01},
%
\begin{equation}\label{Eq.dqdt}
\frac{dQ_{\rm HII}}{dt}= \frac{\dot n(t)}{\langle n_{\rm H} \rangle}-
\frac{\alpha_{\rm B}\chi C \langle n_{\rm H} \rangle Q_{\rm HII}}{a^3(t)} \,\,.
\end{equation} 
%
Here, $\dot n(t)$ is the comoving number density of H~{\sc i} ionizing photons per 
unit time, $\langle n_{\rm H} \rangle=1.87\times10^{-7}$ cm$^{-3}$ is the comoving 
number density of the total hydrogen, C is the clumping factor inside the H~{\sc ii} 
bubbles, $a(t)$ is the scale factor and $\alpha_{\rm B}$ is the case $B$ recombination 
coefficient of hydrogen. In Eq.~(\ref{Eq.dqdt}), the clumping factor is defined as 
$C=\langle n_{\rm HII}^2 \rangle/ \langle n_{\rm H} \rangle ^2$ where $n_{\rm HII}$ 
is the number density of H~{\sc ii}. The solution to the Eq.~(\ref{Eq.dqdt}), 
$Q_{\rm H II}$, at any redshift $z_0$ is given by,
%
\begin{equation}\label{Eq.q}
\begin{aligned}
Q_{\rm H II}(z_0)=\frac{1}{\langle n_{H} \rangle} \int^{\infty}_{z_0}\frac{\dot n(z)}{(1+z)H(z)}\, 
\exp \Bigg[-\alpha_{\rm B} \langle n_{H} \rangle \\
\times \int^{z}_{z_0}{dz'\frac{\chi(z') C(z')(1+z')^2}{H(z')}}\Bigg] dz\,\,.
\end{aligned}
\end{equation} 
%
The reionization is completed when $Q_{\rm HII}(z_{\rm re})$ 
becomes unity and the redshift $z_{\rm re}$ is called as the redshift of reionization.

We need the quantities $\dot n(z)$ and $C(z)$ to evaluate the evolution of 
$Q_{\rm H II}(z)$. Note that the contribution to $C(z)$ comes only from the ionized 
regions, and hence it is important to estimate the distribution of the self-shielded 
regions. This implies that $C(z)$ depends on $\Gamma_{\rm HI}(z)$. In fact, it is 
straightforward to obtain $C$ from simulation boxes for an assumed value of 
$\Gamma_{\rm HI}$. As described earlier, $\Gamma_{\rm HI}(z)$ depends on 
$f(N_{\rm HI}, z)$ and the comoving emissivity $\epsilon_{\nu}(z)$, which in turn 
determines the $\dot n(z)$ through the relation
%
\begin{equation}\label{Eq.ndot}
\dot n(z) = \int^{\infty}_{\nu_{912}}{\frac{\epsilon_{\nu} (z) d\nu}{h \nu}}\,\,.
\end{equation} 
%
Therefore, the values of $\dot n(z)$ and $C(z)$ cannot be chosen independently, 
they are rather related to each other through $\Gamma_{\rm HI}$ (or equivalently 
on $\epsilon_{\nu}$). In this paper, we obtain the $Q_{\rm HII}$ for a given 
$\epsilon_{\nu}(z)$ by self-consistently estimating the three quantities $\dot n(z)$, 
the UVB within H~{\sc ii} bubbles and $C(z)$.

The $\epsilon_{\nu}(z)$ has $f_{\rm esc}$ as a free parameter that we constrain using 
measurements of mean H~{\sc i} fraction and the $\tau_{\rm el}$ from Planck. Using 
the model predictions of $Q_{\rm HII}(z)$, the $\tau_{\rm el}$ is given by
%
\begin{equation}\label{Eq.taueff1}
\tau_{\rm el}(z)=c\,{\langle n_{H}}\rangle \, \sigma_{\rm T} \int^{z}_{0} \chi(z') 
\,Q_{\rm HII}(z')\,\frac{(1+z')^2}{H(z')} \,dz' \,,
\end{equation}
%
where, $ \sigma_{\rm T}$ is the Thompson electron scattering cross-section. We take 
$ \chi(z)=1.16$ for $z<4$ where helium is assumed to be predominantly in He~{\sc iii} 
and $ \chi(z)=1.08$ for $z>4$ where helium is mostly in He~{\sc ii}.

In the following section, we review the comoving H~{\sc i} ionizing emissivity 
$\epsilon_{\nu}(z)$ from different sources used in our models.

\section{Ionizing Emissivity}\label{sec3}
%
%
\begin{figure}
\centering
\includegraphics[totalheight=0.31\textheight, trim=0cm 0cm 3.5cm 0cm, clip=true, ]{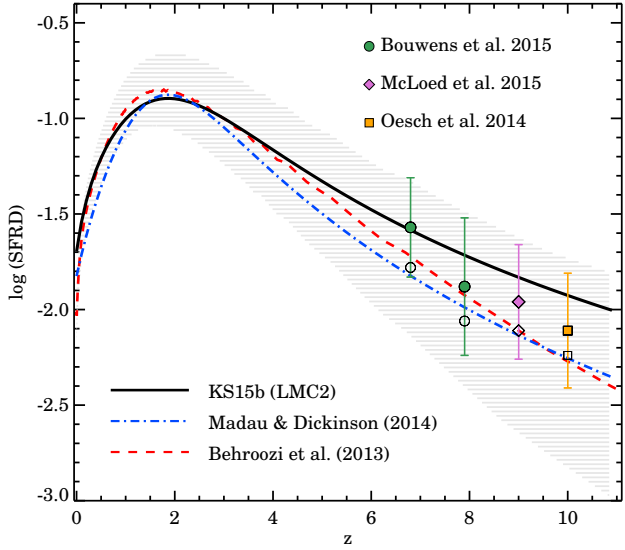}
\caption{The SFRD$(z)$ in units  ${\rm M_{\odot}\, yr^{-1}\, Mpc^{-3}}$ from \citetalias{Khaire15ebl} is shown 
along with the SFRD($z$) obtained by \citet{Madau14} and by \citet[][hatched gray region shows 
1-$\sigma$ range]{Behroozi13} scaled to match the Salpeter IMF. The data points are the SFRD$(z)$
calculated using the respective luminosity functions down to $L_{\rm min}$ =0.01$L^*$. Empty 
symbols are obtained with no dust corrections and filled symbols are obtained using the dust 
corrections from \citetalias{Khaire15ebl}. See text for more details.}
\label{fig2}
\end{figure}
%
%
The ionizing emissivity is estimated by considering the sources of H~{\sc i} ionizing photons 
($\lambda \le 912$~\AA) are only QSOs and star-forming galaxies. Therefore, the total comoving 
emissivity, $\epsilon_{\nu}(z)$, is the sum of the comoving QSO emissivity, $\epsilon^Q_{\nu}(z)$, 
and the comoving galaxy emissivity, $\epsilon^G_{\nu}(z)$. For the power-law spectral energy 
distribution (SED; for $\lambda \le 912$~\AA) it can be written as,
%
\begin{equation}\label{Eq.eps}
\epsilon_{\nu}(z) = \Big(\frac{\nu}{\nu_{912}}\Big)^{\alpha} \epsilon^Q_{\nu_{912}}(z)+ 
\Big(\frac{\nu}{\nu_{912}}\Big)^{\beta}\epsilon^G_{\nu_{912}}(z)\,\,.
\end{equation} 
%
Here, $\epsilon^Q_{\nu_{912}}(z)$ and $\epsilon^G_{\nu_{912}}(z)$ are emissivities at 912~\AA, 
and $\alpha$ and $\beta$ are the indices of power-law SEDs for QSOs and galaxies, respectively, 
as given below.

The $\epsilon^Q_{\nu_{912}}(z)$ in units of ${\rm erg\, s^{-1}\, Hz^{-1} Mpc^{-3}}$ is taken 
from \citetalias{Khaire15puc}. It has been obtained by fitting various QLFs\footnote{See, 
Table 1 and Section 3.1 of \citetalias{Khaire15puc}.} 
\citep{Schulze09, Croom09, Glikman11, Masters12, Ross13, Palanque13, McGreer13, Kashikawa15} 
and integrating the contribution of QSOs down to a luminosity of 0.01$L^*$, where $L^*$ is
the characteristic break luminosity of a QLF. It is given by,   
%
\begin{equation}\label{Eq.Eqso}
\epsilon^Q_{\nu_{912}}(z)=10^{24.6}\,(1+z)^{5.9}\,\frac{\exp(-0.36z)}{\exp(2.2z) + 25.1}\,\,,
\end{equation}
%
with the $\alpha=-1.4$ for $\lambda < 912$~\AA~in Eq.~(\ref{Eq.eps}) \citep{Stevans14}. 
Later, in this paper we also use the $\epsilon^Q_{\nu_{912}}(z)$ obtained using the QLFs 
by \citet{Giallongo15} at $z>4$ to explore the models of H~{\sc i}
reionization with only QSOs.

We use the $\epsilon^G_{\nu_{912}}(z)$ from \citetalias{Khaire15ebl}. In \citetalias{Khaire15ebl},
for different extinction curves, we determine self-consistent combinations of star formation 
rate density, SFRD($z$), and dust attenuation magnitudes, $A_{\rm FUV}(z)$, at the far-ultraviolet (FUV) band 
(central $\lambda=1500$~\AA) using various multi-wavelength and multi-epoch galaxy luminosity 
functions. We find that the combination of SFRD($z$) and $A_{\rm FUV}(z)$ obtained for an 
average extinction curve ($k_{\nu}$) of the Large Magellanic Cloud Supershell 
\citep[LMC2;][]{Gordon03} reproduces various observations. We use the combination of SFRD($z$) 
(in units ${\rm M_{\odot}\, yr^{-1}\, Mpc^{-3}}$) and $A_{\rm FUV}(z)$ (in magnitudes) obtained 
for the LMC2 extinction curve in \citetalias{Khaire15ebl} that is given by,
%
\begin{equation}\label{Eq.sfrd}
{\rm SFRD}(z)=\frac{a+bz}{1+(z/c)^d}\,\, ;  \,\, A_{\rm FUV}(z)=\frac{a_0+b_0z}{1+(z/c_0)^{d_0}}\,\,,
\end{equation}
%
with $a=2.01\times10^{-2}$, $b=8.48\times10^{-2}$, $c=2.5$, $d=3.09$, $a_0=1.42$, $b_0=0.93$, 
$c_0=2.08$ and $d_0=2.2$. We use these to compute the $\epsilon^G_{\nu}(z)$ as,
%
\begin{equation}\label{Eq.conv}
\epsilon^G_{\nu} (z)=C_{\nu}(z)\int_{z}^{\infty}\frac{{\rm SFRD}(z')
\,\,l_{\nu}[t(z)-t(z'), Z]\,\,dz'}{(1+z')H(z')}\, ,
\end{equation}
%
with the dust correction $C_{\nu}(z)$ at $\lambda>912$~\AA,
%
\begin{equation}\label{Eq.conv}
C_{\nu}(z)= 10^{-0.4\,A_{\rm FUV}(z)\frac{k_{\nu}}{k_{\rm FUV}}}.
\end{equation}
%
For $\lambda \le 912$~\AA, we take $C_{\nu}(z)=f_{\rm esc}(z)$ assuming that the SED does 
not get modified by the dust at $\lambda<912$~\AA. This can be interpreted as if the H~{\sc i} 
ionizing photons escape through the holes in the galaxies \citep{Fujita03, Paardekooper11} or 
generated by few unobscured sources or runaway stars \citep{Gnedin08, Conroy12}. The free 
parameter $f_{\rm esc}(z)$  is a luminosity weighted angle averaged escape fraction treated 
in the same way as given in \citetalias{HM12}. For $\lambda < 228$~\AA, we take $C_{\nu}(z)=0$ 
assuming that He~{\sc ii} ionizing photons do not escape from galaxies.
The $l_{\nu}[t(z)-t(z'), Z]$ is the specific luminosity of a simple stellar population 
(in units of erg s$^{-1}$ Hz$^{-1}$ per unit total mass of stars formed) at redshift $z$ having 
the age of $t_0=t(z)-t(z')$ that went through an instantaneous burst of star formation at 
redshift $z'$ with an average metallicity $Z$. We obtain this using a population synthesis 
code {\sc starburst99} \citep{Leitherer99} for a \citet{Salpeter55} initial mass function (IMF) 
and a constant metallicity of 0.4 times solar value ($Z_{\odot}$) for simplicity (i.e, $Z=0.008$). 
We direct readers to Section 8 of \citetalias{Khaire15ebl} for discussions on the effect of 
using different model parameters of stellar population on the derived SFRD($z$) and $A_{\rm FUV}(z)$. 
We obtain a simple fit to $\epsilon^G_{\nu}(z)$ at 
912~\AA~ in units of ${\rm erg\, s^{-1}\, Hz^{-1} Mpc^{-3}}$ as
%
\begin{equation}\label{Eq.galemis}
\epsilon^G_{\nu_{912}}(z)=f_{\rm esc}(z) \times 10^{25} \frac{3.02+13.12z}{1 +(z/2.44)^{3.02}}\,.
\end{equation}
At $\lambda <912$~\AA, we find that $\beta=-1.8$ in Eq.~(\ref{Eq.eps}) 
approximates the SED for the assumed stellar population model. We verify that this
exponent reproduces the $\Gamma_{\rm HI}$ and $\dot n$ generated by the intrinsic spectrum.
Note that the exponent $\beta$ can have slightly different value depending on the assumed 
metallicity and the IMF \citep[see,][]{Becker13}.   

We use a fixed IMF and metallicity to estimate $\epsilon^G_{\nu_{912}}(z)$.
However, using different IMF and metallicities can change the intrinsic ionizing emissivity 
(i.e $\epsilon^G_{\nu_{912}}/f_{\rm esc}$) generated inside galaxies. 
We estimate the uncertainty in this emissivity,
while consistently reproducing the same $\epsilon_{\rm FUV}(z)$ used to 
determine the SFRD($z$), for different IMFs and metallicities. We find that, 
when we change metallicity from $Z=0.4 Z_{\odot}$ to $Z=0.005Z_{\odot}$,
the change in the intrinsic ionizing emissivity is
less than $7\%$. Instead of our fiducial Salpeter IMF, when we use the \citet{Kroupa01} 
IMF (with exponents 1.3 and 2.3 for mass ranges 0.1 to 0.5 M$_{\odot}$ and 0.5 to 
100 M$_{\odot}$, respectively), increase in the intrinsic ionizing emissivity is less than 
$13\%$. However, using top heavy IMFs or the rotation in stars \citep{Topping15} 
the intrinsic ionizing emissivity can be increased significantly.  

In Fig.~\ref{fig2}, we show the SFRD($z$) used here (from Eq.~\ref{Eq.sfrd}). For comparison, 
we also show the SFRD($z$) obtained by \citet{Madau14} and \citet{Behroozi13}. The SFRD($z$) from 
the latter is multiplied by 1.7 to match the result for the Salpeter IMF used in other estimates. 
At $z>7$, our SFRD($z$) is higher than the mean SFRD($z$) given in these papers by 0.2 to 0.4 
dex (see Fig.~\ref{fig2}). This difference arises because of the differences in the applied dust 
correction and the minimum luminosity ($L_{\rm min}$) down to which the FUV luminosity functions 
are integrated. In Fig.~\ref{fig2}, we show the SFRD obtained using the recent high-$z$ 
measurements of the luminosity functions \citep{Oesch14, Bouwens15, McLeod15}. These points are 
obtained using the same scaling relation used in \citetalias{Khaire15ebl},
${\rm SFRD}(z)=1.25\times 10^{-28}\,\epsilon_{\rm FUV} (z) 10^{0.4 A_{\rm FUV}(z)}$ 
${\rm M_{\odot}\, yr^{-1}\, Mpc^{-3}}$, where, $\epsilon_{\rm FUV}$ is the emissivity at FUV 
band obtained by integrating the FUV luminosity functions down 
to $L_{\rm min}=0.01L^*$. The SFRD($z$) used here is consistent 
with these measurements and is within $1-\sigma$ range predicted by \citet{Behroozi13} 
(see the hatched region in Fig.~\ref{fig2}).
 
For simplicity, we neglect the  diffuse emission processes owing to their negligible contribution 
to UVB such as the He~{\sc ii} Lyman and Balmer continuum emission, He~{\sc ii} Lyman-$\alpha$ 
line emission and He~{\sc i} Lyman continuum emission. 

Using the $\epsilon_{\nu}(z)$ explained here, we place constraints on $f_{\rm esc}(z)$ at 
different epochs as discussed in the following section. The resulting trend in $f_{\rm esc}(z)$ 
prompted us to speculate on different QSO emissivities at $z>3.5$ which are also presented later. 
\section{Results and discussions}\label{sec4}
The Lyman-$\alpha$ absorption seen in the spectra of high-$z$ QSOs and the diminishing population 
of Lyman-$\alpha$ emitters suggest that the redshift $z_{\rm re}\sim6$ is the most likely epoch 
where the process of H~{\sc i} reionization is completed. Therefore, we roughly assume that 
the post-reionization era to be $z\le 6$ and pre-reionization era to be $z\ge 6$. However, we 
consider the models of H~{\sc i} reionization where a very late reionization is allowed
i.e, $z_{\rm re}\ge 5.5$. The results are discussed in the following sections.
\subsection{Escape fraction in the post-reionization era }\label{sec4.1}
%
\begin{figure*}
\centering
\includegraphics[totalheight=0.75\textheight, trim=7.5cm 0cm 0cm 0cm, clip=true, angle=90]{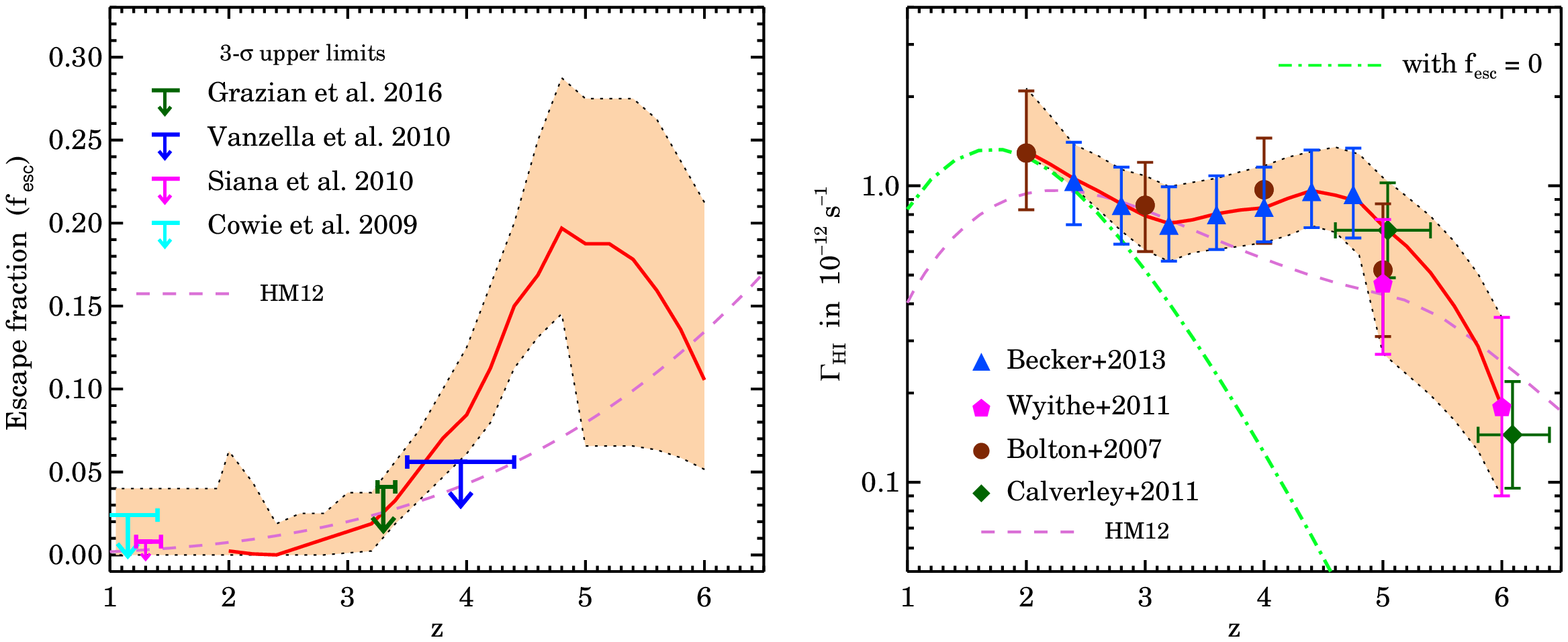}
\caption{The derived constraints on the $f_{\rm esc}(z)$ (\emph{left-hand panel}) from the measurements 
of $\Gamma_{\rm HI}(z)$ (\emph{right-hand panel}). The $\Gamma_{\rm HI}(z)$ measurements are taken from 
\citet{Becker13, Bolton07, Calverley11, Wyithe11}. \emph {Green dot-dash curve} in right panel shows $\Gamma_{\rm HI}(z)$
obtained using only $\epsilon^Q_{\rm 912}(z)$ given in Eq.~(\ref{Eq.Eqso})
i.e with $f_{\rm esc}=0$. Shaded region in both panels show the range in  
the $f_{\rm esc}(z)$ (on \emph{left}) and corresponding $\Gamma_{\rm HI}(z)$ (on \emph{right}). 
Solid red line shows mean $f_{\rm esc}(z)$ that reproduces the mean $\Gamma_{\rm HI}(z)$ measurements. 
We also show 3-$\sigma$ upper limits on mean $f_{\rm esc}(z)$ from various observations 
\citep{Cowie09, Siana10, Vanzella10esc, Grazian15}. The range of $f_{\rm esc}(z)$ obtained here 
is consistent with these upper limits at $z<3.5$. However $f_{\rm esc}(z)$ needs a steep rise from 
$z\sim3$ to $z\sim5$ from $\sim 0.05$ to $\sim 0.2$ to match a nearly constant $\Gamma_{\rm HI}(z)$ 
over this $z$-range. We also show the $f_{\rm esc}(z)$ used by \citetalias{HM12} and 
corresponding $\Gamma_{\rm HI}(z)$. See the text for more details. }
\label{fig3}
\end{figure*}
%
To constrain $f_{\rm esc}$ in the post-reionization era, we use $\epsilon_{\nu}(z)$ as given in 
Eq.~(\ref{Eq.eps}) which has contributions from both QSOs and galaxies. At any redshift $z_0$, the UVB 
depends on $\epsilon_{\nu}$ at $z\ge z_0$. Therefore, we need $f_{\rm esc}$ at $z>z_0$ to estimate 
the UVB at $z_0$. However, at $z>5.5$, the mean free path of  H~{\sc i} ionizing photons 
($\lambda_{\rm mfp}$) is small enough \citep[$\lambda_{\rm mfp}<10$ pMpc;][]{Worseck14LLS} so that 
the UVB is essentially contributed by the local sources. Therefore, we start from redshift $z_0\sim6$
and generate the UVB (and $\Gamma_{\rm HI}$) by taking $f_{\rm esc}(z>z_0)$ as a constant free parameter
(by solving Eq.~(\ref{Eq.uvb}) with $Q_{\rm HII}=1$). The $f_{\rm esc}(z_0)$ is then adjusted to generate
the $\Gamma_{\rm HI}(z_0)$ that matches with the measurement at $z_0$. Once fixed at $z_0\sim6$, 
we use this $f_{\rm esc}(z\ge z_0)$ and calculate the UVB at a lower redshift $z'<z_0$ by taking 
$f_{\rm esc}$ as a constant free parameter in a small redshift interval $z'\le z<z_0$ and fix 
its value to generate the $\Gamma_{\rm HI}(z')$ measurement. This gives the $f_{\rm esc}(z\ge z')$. 
We repeat this procedure for lower $z<z'$ down to $z\sim2$ and obtain the $f_{\rm esc}(z)$ over $2<z<6$. 
This method is identical to the progressive fitting method we used in \citetalias{Khaire15ebl} to 
simultaneously obtain the $A_{\rm FUV}(z)$ and SFRD$(z)$.

To obtain the $f_{\rm esc}(z)$, we used following measurements of $\Gamma_{\rm HI}(z)$. At $2<z<5$, 
we take the $\Gamma_{\rm HI}$ from \citet{Becker13} along with the measurements by \citet{Bolton07}.
These are obtained using the observed mean Lyman-$\alpha$ flux decrement, the opacity of H~{\sc i} 
ionizing photons and measurements of the IGM temperature by \citet{Becker11}. At $z\sim$ 5 and 6, 
we consider the $\Gamma_{\rm HI}$ obtained from the mean Lyman-$\alpha$ opacity by  \citet{Wyithe11} 
and from the QSO proximity effect by \citet{Calverley11}. These measurements are shown in the 
right panel of Fig.~\ref{fig3}. 
We linearly interpolate these measurements over the redshift range $5<z<6$.

The resultant range in the required $f_{\rm esc}(z)$ and corresponding range in the 
$\Gamma_{\rm HI}(z)$ generated from it is shown in the Fig.~\ref{fig3}. At lower redshifts 
$z<2$, the $f_{\rm esc}$ can have values between 0 to 0.04 as demonstrated by us in 
\citetalias{Khaire15puc}. For that, we have used the $\Gamma_{\rm HI}$ inferred by \citet{Kollmeier14} 
and \citet{Shull15} by calibrating their hydrodynamical simulations of the IGM to match the column 
density distribution of low-$z$ IGM reported by \citet{Danforth14} 
\citep[see also,][for different method]{Wakker15}. Including this low-$z$ $f_{\rm esc}$, it is evident 
from  Fig.~\ref{fig3} that for $z<3.5$, the constant $f_{\rm esc}$ of 0 to 0.05 is sufficient 
to generate the measured $\Gamma_{\rm HI}(z)$. This $f_{\rm esc}$ is consistent with the recent 
3-$\sigma$ upper limits on the average $f_{\rm esc}$ obtained by stacking the sample of galaxies 
used in respective studies at $z<0.9$ \citep{Grimes09, Bridge10, Leitet13} and at $0.9<z<3.3$ 
\citep{Cowie09, Siana10, Vanzella10esc, Grazian15} as shown in the left panel of Fig.~\ref{fig3}.
These reported 3-$\sigma$ upper limits are usually given in terms of relative escape fraction 
$f_{\rm esc}^{\rm rel}$ that is related to the absolute escape fraction $f_{\rm esc}$ as following
%
\begin{equation}\label{Eq.fesc0}
f_{\rm esc}^{\rm rel}=\frac{(L_{\rm FUV}/L_{\rm LyC})_{\rm int}}{(L_{\rm FUV}/L_{\rm LyC})_{\rm obs}}\exp(\tau _{\rm IGM})
=f_{\rm esc} 10^{0.4A_{\rm FUV}}\,,
\end{equation}
%
where, $\tau _{\rm IGM}$ is the effective optical depth encountered by Lyman continuum photons 
(LyC, $\lambda \le 912$~\AA) while travelling from the source to the earth due to the IGM. 
We convert these $f_{\rm esc}^{\rm rel}$ into absolute escape fraction $f_{\rm esc}$ by taking 
into account the $A_{\rm FUV}$ used in our galaxy model (see Eq.~\ref{Eq.sfrd}) and the 
difference between the ratio of the intrinsic luminosities $(L_{\rm FUV}/L_{\rm LyC})_{\rm int}$ 
assumed in these references and the one we get 
from our model. 

On the contrary to $z<3.5$, at $3.5<z<4.8$, the required $f_{\rm esc}(z)$ needs a steep rise 
up to $\sim0.15$ to $0.25$ for maintaining the observed trend of a nearly constant $\Gamma_{\rm HI}(z)$. 
The main reason behind this required rapid increase in $f_{\rm esc}$ from $z\sim3.5$ to $z\sim5$ is 
that the galaxies need to produce large number of H~{\sc i} ionizing photons to compensate for the  
rapid decline in the fiducial QSO emissivity by a factor $\sim10$, and to maintain the nearly constant 
$\Gamma_{\rm HI}(z)$ measured in this $z$ range. A similar but less rapid increase in
$f_{\rm esc}\propto (1+z)^{3.4}$ was also required by \citetalias{HM12} as shown in Fig.~\ref{fig3}. 
However, note that the $\Gamma_{\rm HI}(z)$ used by \citetalias{HM12} was from the previous measurements 
of \citet{Becker07} that decreased by a factor of $\sim$2 from $z=3$ to 5 (see, Fig.~\ref{fig3}).

The measured $\Gamma_{\rm HI}$ at $z\sim6$ is a factor of 4 to 10 times smaller than the 
$\Gamma_{\rm HI}$ at $z\sim5$. This decrease in $\Gamma_{\rm HI}$ from $z\sim5$ to $\sim6$ is 
mainly because of the increasing opacity of the IGM \citep{InoueAK14}. We find that the $f_{\rm esc}$ 
of 0.05 to 0.2 is required at $z=6$. It will be interesting to see what values of $f_{\rm esc}$ at 
$z \ge 6$ are allowed by the H~{\sc i} reionization process. We explore this in the following section.
\subsection{Escape fraction in the pre-reionization era}\label{sec4.2}
%
\begin{figure*}
\centering
  \includegraphics[totalheight=0.75\textheight, trim=10.5cm 0cm 0cm 0cm, clip=true, angle=90]{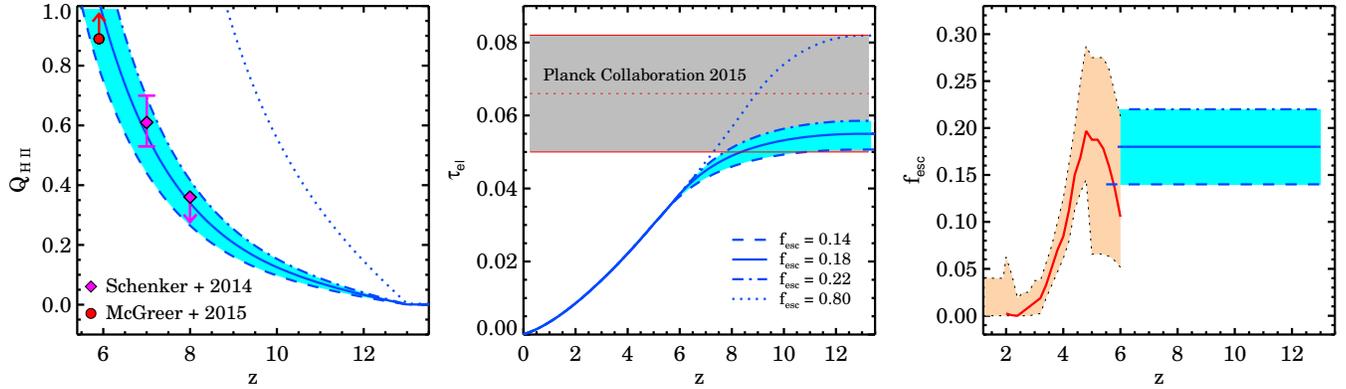}
\caption{The range in $Q_{\rm HII}(z)$ (\emph{left-hand panel}) and the range in corresponding $\tau_{\rm el}(z)$ 
(\emph{central panel}) obtained by using the constant $f_{\rm esc}(z)$ at $z \ge z_{\rm re}$ 
(\emph{right-hand panel; shaded region in cyan}) are shown. We find that the $f_{\rm esc}=0.14$ ($z_{\rm re}=5.5$ 
and $\tau_{\rm el}=0.05$) to $f_{\rm esc}=0.22$ ($z_{\rm re}=6.3$ and $\tau_{\rm el}=0.059$) are consistent 
with the recent $\tau_{\rm el}$ constraints from Planck (\emph{gray shaded} region in \emph{central panel})
and measurements of mean H~{\sc i} fraction (points in the \emph{left-hand panel}) by \citet{Schenker14} and 
\citet{McGreer15}. Maximum allowed $f_{\rm esc}$ only from $\tau_{\rm el}$ is $0.8$ (with $z_{\rm re}=8.8$ 
and $\tau_{\rm el}=0.082$). In the \emph{right-hand panel} we also show the constraints on $f_{\rm esc}(z)$ at 
$z<6$ obtained in Section~\ref{sec4} (see Fig.~\ref{fig3}). Note that, to simultaneously reionize the 
Universe and obtain the $\Gamma_{\rm HI}(z)$ measurements, the $f_{\rm esc}(z)$ needs a rapid increase 
from $\sim$0.05 at $z<3.5$ to at least $\sim$ 0.15 at $z\sim5.5$.}
\label{fig4}
\end{figure*}
%
%
In the pre-reionization era, in absence of any observational constraints or theoretical 
inputs, we consider models of H~{\sc i} reionization with a constant  $f_{\rm esc}$ at 
$z \ge z_{\rm re}$. We extrapolate our fiducial SFRD$(z)$ up to $z\sim13$ and start the 
process of reionization from there.\footnote{We note that, any starting redshift 
$z\ge 13$ has negligible effect on the $z_{\rm re}$ for the models with constant 
$f_{\rm esc}<0.4$.} We find that $\tau_{\rm el}$ measurements of \citet{Planck15} allow 
the minimum constant $f_{\rm esc}$ of 0.14 where the reionization completes at $z_{\rm re}=5.5$ 
and a maximum constant $f_{\rm esc}$ as high as 0.8 where the reionization completes at 
$z_{\rm re}=8.8$. For different $f_{\rm esc}$, we obtain $Q_{\rm HII}(z)$ and evaluate 
$\tau_{\rm el}$. Results for some of these models are shown in Fig.~\ref{fig4}. 
We find that the $Q_{\rm HII}(z)$ obtained for the constant $f_{\rm esc}$ range from 0.14 to 
0.22 satisfies the recently inferred mean H~{\sc i} fraction in the IGM at $z=5.9$  
by \citet{McGreer15} using the dark pixel statistic in high-$z$ QSO spectra and at $z\sim 7$ 
and $z\sim 8$ by \citet{Schenker14} using the observations of the diminishing fraction 
of high-$z$ Lyman-$\alpha$ emitters (see Fig.~\ref{fig4} left panel). As explained earlier, 
this range of $f_{\rm esc}$ is also consistent with the $f_{\rm esc}$ required to generate 
$\Gamma_{\rm HI}$ measurements at $z\sim6$. (See appendix~\ref{app} for the 
discussion on the $\Gamma_{\rm HI}(z)$ and $C(z)$ at $z\ge z_{\rm re}$.)
The recent measurements of $\tau_{\rm el}$ 
from Planck satellite along with the updated SFRD$(z)$ has helped us to reduce the $f_{\rm esc}$ 
as small as $0.14$ \citep[see also][]{Mashian15, Mitra15, Robertson15, Bouwens15reion, Vangioni15} 
from very high values of $f_{\rm esc}$ inferred previously from $\tau_{\rm el}$ reported by
WMAP \citep[such as, \citetalias{HM12},][]{Choudhury05, Kuhlen12, Fontanot14}. We consider 
a model with a constant $f_{\rm esc}=0.18$ as our fiducial model where the reionization 
completes at $z_{\rm re}=5.9$ and $\tau_{\rm el}=0.055$. 

Combining the $f_{\rm esc}$ constraints from $z\le 6$ (as shown in the right panel of 
Fig.~\ref{fig4}), an evolution in $f_{\rm esc}(z)$ of at least a factor 3 from 0.05 to 0.15 
is required from $z\sim3.5$ to $z\sim5.5$ to simultaneously satisfy recent observational 
constraints on H~{\sc i} reionization and $\Gamma_{\rm HI}$ measurements at all $z$. This 
redshift range ($3.5<z<5.5$) corresponds to an elapsed time of $7.5\times 10^8$ yr, 
only 4 times the dynamical time scale at $z=4.5$ \citep{Samui07}. This trend in 
$f_{\rm esc}$ implies that the population of galaxies at $z>3.5$ are different from their 
local counterparts. It may either mean that the high-z galaxies are more porous than 
the low-$z$ ones or the stellar efficiency to generate large amount of H~{\sc i} ionizing 
photons has increased via rapid evolution in IMF, metallicity or the rotations in stars 
\citep{Topping15}. However, in the latter case, these parameters should evolve rapidly to 
provide a higher ratio of $L_{\rm LyC}$ to $L_{\rm FUV}$ (see Eq.~\ref{Eq.fesc0}). 
For example, changing metallicity from 0.4Z$_{\odot}$ to 0.005Z$_{\odot}$ increases the 
above ratio only by $< $10\%. Therefore, a rapid evolution in metallicity alone will not 
provide rapid increase in $f_{\rm esc}$. Many theoretical studies predict increasing 
$f_{\rm esc}$ with $z$ via various mechanisms such as the luminosity dependent $f_{\rm esc}$ 
\citep{Ferrara13}, a large $f_{\rm esc}$ from faint low mass galaxies whose number 
density increases with $z$ \citep{Alvarez12, Cai14, Fontanot14}, the supernovae dominated primordial 
galaxies \citep{Yajima09} and evolution in the covering factors of clumps in the interstellar 
medium \citep{Fernandez11, Roy15}. However, the main difficulty of using these mechanisms 
to explain our results is that the corresponding physical changes in the properties of 
galaxies have to occur over a very short period of time.

There is a growing evidence of $f_{\rm esc}$ being low even at high $z$. The earlier 
reported measurements of high $f_{\rm esc}$ values \citep[such as,][]{Steidel01, Shapley06} 
are found to be dominated by the contamination from low-$z$ interlopers 
\citep{Siana15, Mostardi15, Grazian15}. In addition, simulations of \citet{Ma15} show 
that the time average value of $f_{\rm esc}$ is $\le 0.05$ without any strong dependence 
on the properties of galaxies and redshift. In light of these,  we now explore the models 
with low $f_{\rm esc}$ of H~{\sc i} ionizing photons from galaxies. This will require an 
enhanced contribution from QSOs to H~{\sc i} ionizing photons at $z>3.5$. The QSO emissivity 
at $z>3.5$ is fairly uncertain because of the ill-constrained faint end slope and the 
characteristic turnover luminosity $L^*$ of the QLFs. This allows us some room for the 
exploration. 
\subsection{Prediction for optimal QSO model: $\epsilon^Q_{\rm 912}$(z)}\label{sec4.3}
%
\begin{figure*}
\centering
\includegraphics[totalheight=0.75\textheight, trim=7.5cm 0cm 0cm 0cm, clip=true, angle=90]{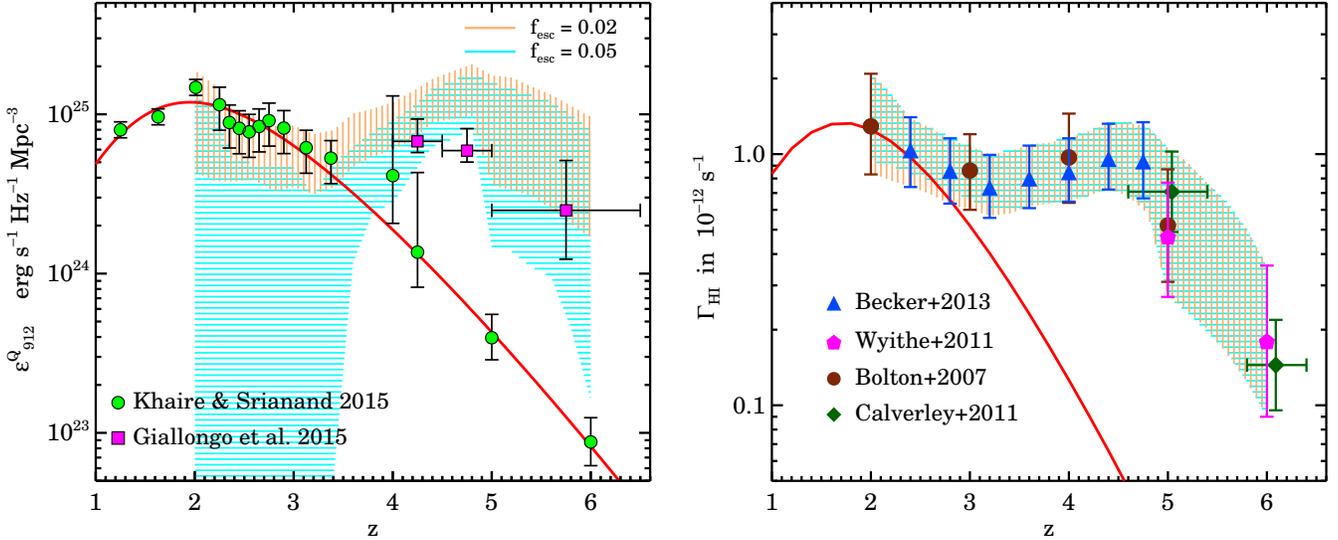}
\caption{Left-hand panel: Constraints on the $\epsilon^Q_{\rm 912}(z)$ at $z>2$ obtained assuming a 
constant $f_{\rm esc}=0.02$ and $0.05$ of H~{\sc i} ionizing photons from galaxies. The required 
range of $\epsilon^Q_{\rm 912}(z)$ when  $f_{\rm esc}=0.05$ is shown by horizontal stripes. 
Notice at 
$2<z<3.5$ the $f_{\rm esc}=0.05$ is sufficient to generate $\Gamma_{\rm HI}$ measurements down to
its 1$-\sigma$ low value purely from galaxies and no additional contribution is needed from QSOs. 
Allowed range of $\epsilon^Q_{\rm 912}(z)$ when  $f_{\rm esc}=0.02$ is shown by vertical stripes. 
The \emph{green circles} are from compilation by \citetalias{Khaire15puc} and \emph{red curve} is a 
fit through them as given in Eq.~(\ref{Eq.Eqso}). \emph{Magenta squares} are from QLFs of \citet{Giallongo15}. 
Right-hand Panel: The corresponding $\Gamma_{\rm HI}(z)$ measurements are shown (see Fig.~\ref{fig3}). 
The description of horizontal and vertical stripes
is same as given for the left panel. \emph{Red curve} shows $\Gamma_{\rm HI}$
computed using $\epsilon^Q_{\rm 912}(z)$ fit shown in the left panel with $f_{\rm esc}=0$. }
\label{fig5}
\end{figure*}
%
\citet{Giallongo15} discovered 22 faint active galactic nuclei (AGN) candidates at $4<z<6$ through 
multi-wavelength observations of high-$z$ galaxies. The QLFs derived using these 
AGN have $L^*$ an order of magnitude smaller and number density $\phi^*$ at least 
two orders of magnitude higher as compared to other QLFs 
\citep[such as,][]{Masters12, McGreer15, Kashikawa15}. This eventually provides an 
emissivity of the H~{\sc i} ionizing photons that is $\sim$10 to $\sim$50 times higher 
than the other QLFs, assuming the usual escape fraction of unity and the same SEDs that 
one takes for the bright QSOs. Interestingly, as shown by \citet{Giallongo15} the 
emissivity obtained in this way can generate the $\Gamma_{\rm HI}(z)$ measurements without 
requiring any galaxy contribution at $4<z<6$. Motivated by these QLF measurements, 
we take the  H~{\sc i} ionizing QSO emissivity, $\epsilon^Q_{\rm 912}(z)$, as a free 
parameter at $2<z<6$ and constrain its values using $\Gamma_{\rm HI}(z)$ measurements by 
fixing the $f_{\rm esc}$ from galaxies to be in the range of 0.02 to 0.05. 

We follow the same method described in Section~\ref{sec4.1}, where instead of $f_{\rm esc}$ 
now we find $\epsilon^Q_{\rm 912}(z)$ consistent with the $\Gamma_{\rm HI}$ measurements 
by assuming a constant $f_{\rm esc}$ of 0.02 and 0.05. The results of this exercise are 
shown in Fig.~\ref{fig5} along with the $\epsilon^Q_{\rm 912}(z)$ used here (Eq.~\ref{Eq.Eqso}) 
from compilation of \citetalias{Khaire15puc} and from the QLF of \citet{Giallongo15}. For a 
constant $f_{\rm esc}=0.05$, at $z<3.5$ the minimum required $\epsilon^Q_{\rm 912}(z)$ goes to 
zero since galaxies are sufficient to generate the measured $\Gamma_{\rm HI}$ down to its 
1$-\sigma$ value (see the horizontal striped region in Fig.~\ref{fig5}). However, at $z>3.5$ the 
required $\epsilon^Q_{\rm 912}(z)$ is higher suggesting a need for an increase in the fiducial 
$\epsilon^Q_{\rm 912}(z)$ by at least a factor of $\sim$2 to $\sim$8. For $f_{\rm esc}=0.02$, 
there are more stringent constraints on the $\epsilon^Q_{\rm 912}(z)$. At $z<3.5$ our fiducial 
$\epsilon^Q_{\rm 912}(z)$ is consistent with the allowed range  (see the vertical striped region 
in Fig.~\ref{fig5}). However, at $z>3.5$, there is a need of at least a factor of $\sim$4 
to $\sim$20 rise in our fiducial $\epsilon^Q_{\rm 912}(z)$. Note that, At $z>4$, the QLFs used 
to obtain the $\epsilon^Q_{\rm 912}(z)$ in  \citetalias{Khaire15puc} are taken from 
\citet{Glikman11, Masters12, McGreer13, Kashikawa15}, where the faint end slopes have large 
uncertainties. The $\epsilon^Q_{\rm 912}(z)$ from  \citet{Glikman11} at $z\sim4$ and 
\citet{Giallongo15} at $z\ge4$ satisfy the allowed range when we assume $f_{\rm esc}$ from 
0.02 to 0.05. 
\subsection{H~{\sc i} reionization with only QSOs}\label{sec4.4}
%
\begin{figure*}
\centering
\includegraphics[totalheight=0.75\textheight, trim=10.5cm 0cm 0cm 0cm, clip=true, angle=90]{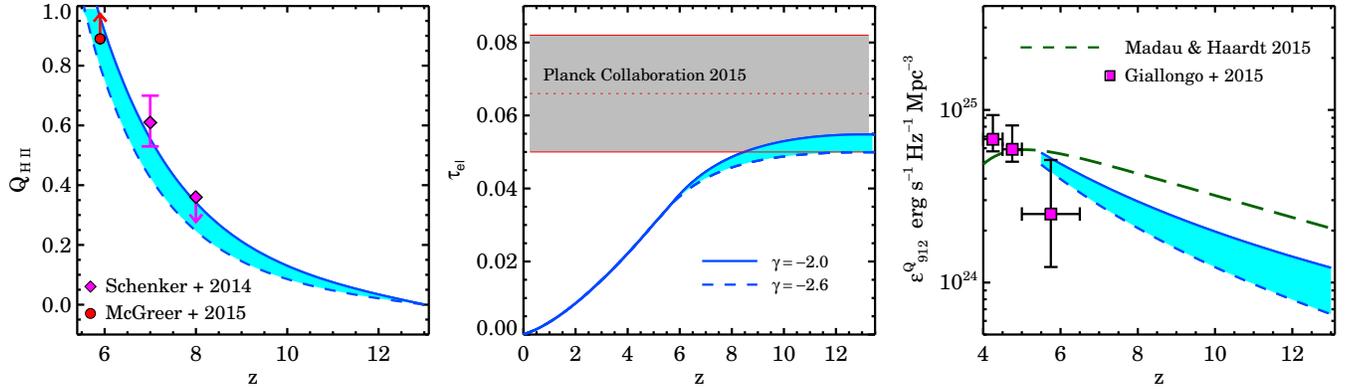}
\caption{The range in $Q_{\rm HII}(z)$ (\emph{left-hand panel}) and the range in corresponding 
$\tau_{\rm el} (z)$  (\emph{central panel}) obtained by using emissivity of QSOs (\emph{right-hand panel; 
shaded region in cyan}) with no contribution from galaxies (i.e. $f_{\rm esc}=0$). We parametrize 
the QSO emissivity at 912~\AA~ with a power-law as $\epsilon^Q_{\rm 912}(z)=9.55\times10^{24}[(1+z)/5]^{\gamma}$ 
${\rm erg\, s^{-1}\, Hz^{-1} Mpc^{-3}}$ and find that the $\gamma$ can have values $-2$ 
(\emph{solid line, right-hand panel}) to $-2.6$ (\emph{dash line, right-hand panel}) that is consistent with 
the QLFs by \citet{Giallongo15} (\emph{magenta squares; right-hand panel}), $\tau_{\rm el}$ constraints 
from Planck (\emph{gray shade; central panel}) and the mean H~{\sc i} fraction by \citet{Schenker14} 
and \citet{McGreer15}  (\emph{points in left-hand panel}). For comparison, we also show the 
$\epsilon^Q_{\rm 912}$ emissivity used by \citetalias{Madau15} (\emph{green long dash; right-hand panel}). 
$\epsilon^Q_{\rm 912}(z)$ used by us is lower than them. See the text for more details.}
\label{fig6}
\end{figure*}
%
The population of faint AGN detected in the recent multi-wavelength surveys 
\citep{Glikman11, Fiore12, Civano11} hints towards the possibility that the QSOs may be 
dominant sources of reionization \citep{Giallongo12}. In light of these, we explore a 
possibility where only QSOs reionize the Universe. We consider a range of QSO emissivities 
at $z>z_{\rm re}$ which can reionize the Universe and at the same time remaining consistent 
with the emissivity of \citet{Giallongo15} and $\tau_{\rm el}$ measurements from \citet{Planck15}.  
We parametrize the range in QSO emissivity at $z>4$ as a power law 
$\epsilon^Q_{\rm 912}(z)=9.55\times10^{24}[(1+z)/5]^{\gamma}$ 
${\rm erg\, s^{-1}\, Hz^{-1} Mpc^{-3}}$ and find that the $\gamma$ can have values
$-2$ to $-2.6$. For this range in $\epsilon^Q_{\rm 912}$, the obtained $Q_{\rm H II}(z)$ 
evolution and the  $\tau_{\rm el}(z)$ are shown in the Fig.~\ref{fig6}. Similar to the 
previous model, the resultant $Q_{\rm H II}(z)$ is also consistent with the mean H~{\sc i} 
fraction measurements of \citet{Schenker14} and \citet{McGreer15}. For $\gamma=-2.0$ (respectively $-2.6$), 
the H~{\sc i} reionization completes at $z_{\rm re}=5.8$ (respectively $5.5$) and it provides 
$\tau_{\rm el}=0.055$ (respectively $0.05$). 

Similar calculations were recently presented by \citet{Madau15} (hereafter \citetalias{Madau15}), 
where they obtain the $z_{\rm re}\sim 5.7$ with $\tau_{\rm el}=0.056$. A small difference between 
our and their models arises from the differences in the SEDs ($\alpha$, See Eq.~\ref{Eq.eps}) 
of QSOs and the clumping factor used. We take $\alpha=-1.4$ following \citet{Stevans14} which 
gives 20\% higher $\dot n$ as compared to the one obtained by \citetalias{Madau15} with 
$\alpha=-1.7$ following \citet{Lusso15}. The differences in clumping factors are shown in appendix~\ref{app}. 
Following the exact prescription given by \citetalias{Madau15} to calculate the contribution 
of QSOs at $z>5$ to the unresolved X-ray background, we find that the $J_{\nu}$ at 2 keV 
is $1.32\times10^{-27}$ (respectively $1.04\times10^{-27}$)
${\rm erg\, cm^{-2}\, s^{-1}\, Hz^{-1} Sr^{-1}}$ for our emissivity models with 
$\gamma=-2.0$ (respectively $-2.6$) that is nearly 45\% (respectively 35\%) of the total X-ray 
background derived by \citet{Moretti12}. Therefore, the emissivity of QSOs taken by us does 
not overproduce the unresolved X-ray background. It is still valid, even if we take the 
factor $R_{\rm II}$, the correction factor to estimate the obscured fraction of AGN, equal 
to 4 \citep[see for e.g.][]{Haardt15, Merloni14} instead of $2$ as taken by \citetalias{Madau15}.

Till now, following \citet{Giallongo15} and \citetalias{Madau15}, we have used a 
luminosity independent SED ($\alpha=-1.4$) to estimate the emissivities from QLF. However,
when faint AGN are included, the QSO luminosity extends over a wide range and 
therefore any correlation between the QSO luminosity and SED may become important. 
We provide a simple way to estimate the luminosity-dependent $\alpha$ via a tight 
correlation found between the optical luminosity at 2500~\AA~(i.e $L_{\nu 2500}$) and the optical 
to X-ray (2 keV) spectral index, $\alpha_{\rm ox}$. This is usually written  
as $\alpha_{\rm ox}=AL_{\nu 2500} -B$, where $A=0.154$ and $B=-3.176$ as reported by 
\citet{Lusso10}, and $A=0.065$ and  $B=-0.509$ as reported by \citet{Stalin10}.
We have used a piece-wise SED to obtain the ionizing emissivity from QLFs 
\citepalias[in][]{Khaire15puc} by taking $L_{\nu} \propto \nu^{\alpha_0}$ with 
$\alpha_0=-0.5$ at $\lambda>2000$~\AA, $\alpha_0=-0.8$ at $2000>\lambda>1000$~\AA~and 
$\alpha_0=\alpha$ at $1000>\lambda>24.8$~\AA~\citep{Stevans14}, and $\alpha_0=-0.9$ at 
$\lambda<24.8$~\AA~\citep{Nandra94}. This coupled with $\alpha_{\rm ox}$ to give a consistent ratio 
of luminosities at 2500~\AA~and 2 keV results into $\alpha=1.1\alpha_{\rm ox}$. This relation 
along with the correlation between $\alpha_{\rm ox}$ and $L_{\nu 2500}$ provides a luminosity 
dependent $\alpha$. We use this and estimate the $\epsilon^Q_{\nu_{912}}$ from the QLF of \citet{Giallongo15}. 
This gives a maximum of 30 (respectively 20)\% lower $\dot n$ and 16 (respectively 13)\% lower
$\Gamma_{\rm HI}$ when we use $\alpha_{\rm ox}$ and $L_{\nu 2500}$ correlation from \citet{Stalin10}
\citep[respectively][]{Lusso10}. This change in $\dot n$ and $\Gamma_{\rm HI}$ allows $\gamma$ to have 
minimum value of $-2.5$ instead of $-2.6$. 

The model presented here relies on the QLFs reported by \citet{Giallongo15}. 
However, a similar study performed by \citet{Weigel15} and \citet{Georgakakis15} 
could not confirm the detection of large number of faint AGN. \citet{Georgakakis15} 
estimated X-ray QLFs at $3<z<5$ and show that the upper limits on the H~{\sc i} 
ionizing emissivity from their QLFs can not generate the measured
$\Gamma_{\rm HI}$ at $z>4$. More observations of similar nature and tight 
constraints on the SEDs of the faint AGN along with their escape fraction of 
UV photons are required to confirm the results of \citet{Giallongo15}.
 
\section{Summary}\label{sec5}
We study the contribution of QSOs and galaxies to the H~{\sc i} ionizing emissivity 
as a function of $z$ that can reionize the Universe satisfying the $\tau_{\rm el}$ 
constraints from the recent CMB measurements and consistently reproduce the 
$\Gamma_{\rm HI}$ measurements at $z\le6$. For this, we use a radiative transfer 
code developed by us to estimate the UV background with the updated QSO emissivity 
\citepalias{Khaire15puc} and star formation rate density \citepalias{Khaire15ebl}.
We use updated H~{\sc i} column density distribution from \citet{InoueAK14} at $z<6$
and the one computed using hydrodynamical simulations in the pre-reionization era.

We constrain the escape fraction ($f_{\rm esc}$) of H~{\sc i} ionizing 
photons from galaxies as a function of $z$. We find a constant 
$f_{\rm esc}=0.14-0.22$ is sufficient to reionize the Universe satisfying the recent 
observational constraints \citep{Planck15, McGreer15, Schenker14}. We show that, a 
constant $f_{\rm esc}\sim 0-0.05$ is sufficient to obtain the measured 
$\Gamma_{\rm HI}$ at $z\le3.5$. However, between $z \sim 3.5$ and $5.5$ 
(i.e over an elapsed time of $\sim 7.5 \times 10^8$ yr) a sharp increase in 
$f_{\rm esc}$ of at least a factor $\sim$3 is required. This is mainly because 
of the rapid decline in the QSO emissivity at $z\ge3$ and the requirement to 
obtain a nearly constant $\Gamma_{\rm HI}(z)$. 

The required rapid increase in $f_{\rm esc}$ may imply either the high-$z$ 
galaxies are more porous than the low-$z$ ones or the stellar efficiency to 
generate large amount of H~{\sc i} ionizing photons has increased with a rapid 
evolution in the properties of galaxies. Confirming this by direct measurements 
of increasing $f_{\rm esc}$ or rapidly evolving nebular emission line ratios 
over $3.5<z<5.5$ will favor the galaxy dominated UV background at high-$z$.
However, the main problem with this scenario is that all the physical 
changes have to occur over a very short time-scale ($\sim 4$ times the 
dynamical time scale at $z=4.5$). This may not be naturally obtained in the 
structure formation models \citep[see for e.g.][]{Ma15}. In addition, there is 
a growing evidence of $f_{\rm esc}$ being low  even at high $z$. The earlier 
reported measurements of high $f_{\rm esc}$ values are found to be dominated 
by the contamination from low-$z$ interlopers \citep{Siana15, Mostardi15, Grazian15}. 

We further explore the possibility of QSOs dominating the H~{\sc i} ionization 
throughout the history of the Universe. We estimate the required QSO emissivities 
assuming a constant $f_{\rm esc}=0.02-0.05$. Under this scenario, at $z>3.5$ our 
fiducial QSO emissivity needs to be enhanced by at least a factor of $\sim$4 to $\sim$20 to 
obtain the measured $\Gamma_{\rm HI}$. However, this enhanced emissivity is 
consistent with the QLF obtained by \citet{Giallongo15}. With a simple extrapolation 
of this emissivity to higher redshifts, we show that QSOs alone can reionize the 
Universe satisfying the observational constraints \citepalias[as also shown by][]{Madau15}. 
Note that in this case, the QSO emissivity is dominated by the low luminosity AGN 
with black hole masses in the range 10$^6$-10$^{8}$ M$_\odot$ probably hosted by low 
mass (i.e $10^9$ M$_\odot$) galaxies. However, models of black hole formation at high-$z$
show that the strong stellar feedback may hinder black hole formation in low mass 
galaxies \citep[see for e.g.][]{Dubois15}. Therefore, independent observational 
confirmations of the findings of \citet{Giallongo15} are important
\citep[see][for non-confirmation]{Weigel15, Georgakakis15} to further 
investigate the QSO dominated UV background scenarios at $z>3.5$.

The QSO-dominated H~{\sc i} reionization can also lead to a range of observable 
signatures. The topology of the ionized bubbles will be distinctly different as 
compared to galaxy-dominated models. This will change the predictions of 21 cm 
signals that can be searched with future experiments. This will also affect the 
constraints on reionization obtained from the observed decrease in the density 
of Lyman-$\alpha$ emitters at $z >6$ \citep[such as][]{Mesinger08, Choudhury15}.
The AGN contamination in high-$z$ Lyman-$\alpha$ emitter samples should also 
show an increasing trend with $z$. In principle, the highly ionized regions 
around these numerous faint AGN can be studied using the presence of the 
He~{\sc ii} $\lambda$1640~\AA~emission line. The thermal history of the IGM will 
be affected because of the hard SEDs of QSOs, the additional heating from the 
He~{\sc ii} ionization in the near zones of QSOs \citep{Bolton12, Padmanabhan14} 
and the X-ray heating \citep{Venkatesan01}. The inferred redshift of He~{\sc ii} 
reionization and the interpretation of observed He~{\sc ii} Lyman-$\alpha$ 
effective optical depths \citep{Worseck14} will be significantly different 
\citep[][ \citetalias{Madau15}]{Furlanetto10, Compostella14}.

All these emphasize a need for improved observational constraints and a careful 
modelling of the above-mentioned observables under both galaxy- and QSO-dominated 
UV background scenarios. The improved $\Gamma_{\rm HI}$ measurements with tight 
constraints on IGM temperature and density are also needed to further study the 
relative contribution of QSOs and galaxies to UV background.
\section*{acknowledgement} 
We thank anonymous referee for the useful comments on the manuscript.
We thank V. Springel for providing {\sc p-gadget3} code. The simulations 
were performed using the Perseus cluster of the IUCAA High Performance Computing 
Centre. We thank A. Paranjape for the useful comments on the manuscript.
V. Khaire acknowledges support from the CSIR. 

\newcommand{\pasp}{PASP}
\def\apj{ApJ}
\def\mnras{MNRAS}
\def\aap{A\&A}
\def\apjl{ApJ}
\def\aj{aj}
\def\physrep{PhR}
\def\pre{PhRvE}
\def\apjs{ApJS}
\def\pasa{PASA}
\def\pasj{PASJ}
\def\nat{Nat}
\def\ssr{SSRv}
\def\aapr{AAPR}
\def\araa{ARAA}

{\small \bibliographystyle{mn2e}
\bibliography{vikrambib}}

\begin{thebibliography}{139}
\providecommand{\natexlab}[1]{#1}

\bibitem[{{Alvarez} et~al.(2012){Alvarez}, {Finlator} \& {Trenti}}]{Alvarez12}
{Alvarez} M.~A., {Finlator} K., {Trenti} M., 2012, \apjl, 759, L38

\bibitem[{{Barger} et~al.(2013){Barger}, {Haffner} \&
  {Bland-Hawthorn}}]{Barger13}
{Barger} K.~A., {Haffner} L.~M., {Bland-Hawthorn} J., 2013, \apj, 771, 132

\bibitem[{{Barkana} \& {Loeb}(2001)}]{Barkana01}
{Barkana} R., {Loeb} A., 2001, \physrep, 349, 125

\bibitem[{{Becker} \& {Bolton}(2013)}]{Becker13}
{Becker} G.~D., {Bolton} J.~S., 2013, \mnras, 436, 1023

\bibitem[{{Becker} et~al.(2007){Becker}, {Rauch} \& {Sargent}}]{Becker07}
{Becker} G.~D., {Rauch} M., {Sargent} W.~L.~W., 2007, \apj, 662, 72

\bibitem[{{Becker} et~al.(2011){Becker}, {Bolton}, {Haehnelt} \&
  {Sargent}}]{Becker11}
{Becker} G.~D., {Bolton} J.~S., {Haehnelt} M.~G., {Sargent} W.~L.~W., 2011,
  \mnras, 410, 1096

\bibitem[{{Becker} et~al.(2013){Becker}, {Hewett}, {Worseck} \&
  {Prochaska}}]{Becker13Ly}
{Becker} G.~D., {Hewett} P.~C., {Worseck} G., {Prochaska} J.~X., 2013, \mnras,
  430, 2067

\bibitem[{{Behroozi} et~al.(2013){Behroozi}, {Wechsler} \&
  {Conroy}}]{Behroozi13}
{Behroozi} P.~S., {Wechsler} R.~H., {Conroy} C., 2013, \apj, 770, 57

\bibitem[{{Benson} et~al.(2013){Benson}, {Venkatesan} \& {Shull}}]{Benson13}
{Benson} A., {Venkatesan} A., {Shull} J.~M., 2013, \apj, 770, 76

\bibitem[{{Bolton} \& {Haehnelt}(2007)}]{Bolton07}
{Bolton} J.~S., {Haehnelt} M.~G., 2007, \mnras, 382, 325

\bibitem[{{Bolton} \& {Haehnelt}(2013)}]{Bolton13}
{Bolton} J.~S., {Haehnelt} M.~G., 2013, \mnras, 429, 1695

\bibitem[{{Bolton} et~al.(2012){Bolton}, {Becker}, {Raskutti}, {Wyithe},
  {Haehnelt} \& {Sargent}}]{Bolton12}
{Bolton} J.~S., {Becker} G.~D., {Raskutti} S., {Wyithe} J.~S.~B., {Haehnelt}
  M.~G., {Sargent} W.~L.~W., 2012, \mnras, 419, 2880

\bibitem[{{Bolton} et~al.(2011)}]{Bolton11}
{Bolton} J.~S., {Haehnelt} M.~G., {Warren} S.~J., {Hewett} P.~C., {Mortlock}
  D.~J., {Venemans} B.~P., {McMahon} R.~G., {Simpson} C., 2011, \mnras, 416,
  L70

\bibitem[{{Borthakur} et~al.(2014){Borthakur}, {Heckman}, {Leitherer} \&
  {Overzier}}]{Borthakur14}
{Borthakur} S., {Heckman} T.~M., {Leitherer} C., {Overzier} R.~A., 2014,
  Science, 346, 216

\bibitem[{{Boutsia} et~al.(2011)}]{Boutsia11}
{Boutsia} K. et~al., 2011, \apj, 736, 41

\bibitem[{{Bouwens} et~al.(2015{\natexlab{a}}){Bouwens}, {Illingworth},
  {Oesch}, {Caruana}, {Holwerda}, {Smit} \& {Wilkins}}]{Bouwens15reion}
{Bouwens} R.~J., {Illingworth} G.~D., {Oesch} P.~A., {Caruana} J., {Holwerda}
  B., {Smit} R., {Wilkins} S., 2015{\natexlab{a}}, \apj, 811, 140

\bibitem[{{Bouwens} et~al.(2012)}]{Bouwens12}
{Bouwens} R.~J. et~al., 2012, \apj, 754, 83

\bibitem[{{Bouwens} et~al.(2015{\natexlab{b}})}]{Bouwens15}
{Bouwens} R.~J. et~al., 2015{\natexlab{b}}, \apj, 803, 34

\bibitem[{{Bridge} et~al.(2010)}]{Bridge10}
{Bridge} C.~R. et~al., 2010, \apj, 720, 465

\bibitem[{{Cai} et~al.(2014){Cai}, {Lapi}, {Bressan}, {De Zotti}, {Negrello} \&
  {Danese}}]{Cai14}
{Cai} Z.~Y., {Lapi} A., {Bressan} A., {De Zotti} G., {Negrello} M., {Danese}
  L., 2014, \apj, 785, 65

\bibitem[{{Calverley} et~al.(2011){Calverley}, {Becker}, {Haehnelt} \&
  {Bolton}}]{Calverley11}
{Calverley} A.~P., {Becker} G.~D., {Haehnelt} M.~G., {Bolton} J.~S., 2011,
  \mnras, 412, 2543

\bibitem[{{Cen} \& {Kimm}(2015)}]{Cen15}
{Cen} R., {Kimm} T., 2015, \apjl, 801, L25

\bibitem[{{Choudhury}(2009)}]{Choudhury09}
{Choudhury} T.~R., 2009, Current Science, 97, 841

\bibitem[{{Choudhury} \& {Ferrara}(2005)}]{Choudhury05}
{Choudhury} T.~R., {Ferrara} A., 2005, \mnras, 361, 577

\bibitem[{{Choudhury} et~al.(2015){Choudhury}, {Puchwein}, {Haehnelt} \&
  {Bolton}}]{Choudhury15}
{Choudhury} T.~R., {Puchwein} E., {Haehnelt} M.~G., {Bolton} J.~S., 2015,
  \mnras, 452, 261

\bibitem[{{Civano} et~al.(2011)}]{Civano11}
{Civano} F. et~al., 2011, \apj, 741, 91

\bibitem[{{Compostella} et~al.(2014){Compostella}, {Cantalupo} \&
  {Porciani}}]{Compostella14}
{Compostella} M., {Cantalupo} S., {Porciani} C., 2014, \mnras, 445, 4186

\bibitem[{{Conroy} \& {Kratter}(2012)}]{Conroy12}
{Conroy} C., {Kratter} K.~M., 2012, \apj, 755, 123

\bibitem[{{Cowie} et~al.(2009){Cowie}, {Barger} \& {Trouille}}]{Cowie09}
{Cowie} L.~L., {Barger} A.~J., {Trouille} L., 2009, \apj, 692, 1476

\bibitem[{{Croom} et~al.(2009)}]{Croom09}
{Croom} S.~M. et~al., 2009, \mnras, 399, 1755

\bibitem[{{Danforth} et~al.(2016)}]{Danforth14}
{Danforth} C.~W. et~al., 2016, \apj, 817, 111

\bibitem[{{Dove} \& {Shull}(1994)}]{Dove94}
{Dove} J.~B., {Shull} J.~M., 1994, \apj, 423, 196

\bibitem[{{Dubois} et~al.(2015){Dubois}, {Volonteri}, {Silk}, {Devriendt},
  {Slyz} \& {Teyssier}}]{Dubois15}
{Dubois} Y., {Volonteri} M., {Silk} J., {Devriendt} J., {Slyz} A., {Teyssier}
  R., 2015, \mnras, 452, 1502

\bibitem[{{Fan} et~al.(2006)}]{Fan06}
{Fan} X. et~al., 2006, \aj, 132, 117

\bibitem[{{Fardal} et~al.(1998){Fardal}, {Giroux} \& {Shull}}]{Fardal98}
{Fardal} M.~A., {Giroux} M.~L., {Shull} J.~M., 1998, \aj, 115, 2206

\bibitem[{{Faucher-Gigu{\`e}re} et~al.(2009){Faucher-Gigu{\`e}re}, {Lidz},
  {Zaldarriaga} \& {Hernquist}}]{FG09}
{Faucher-Gigu{\`e}re} C.~A., {Lidz} A., {Zaldarriaga} M., {Hernquist} L., 2009,
  \apj, 703, 1416

\bibitem[{{Fernandez} \& {Shull}(2011)}]{Fernandez11}
{Fernandez} E.~R., {Shull} J.~M., 2011, \apj, 731, 20

\bibitem[{{Ferrara} \& {Loeb}(2013)}]{Ferrara13}
{Ferrara} A., {Loeb} A., 2013, \mnras, 431, 2826

\bibitem[{{Finkelstein} et~al.(2015)}]{Finkelstein15}
{Finkelstein} S.~L. et~al., 2015, \apj, 810, 71

\bibitem[{{Fiore} et~al.(2012)}]{Fiore12}
{Fiore} F. et~al., 2012, \aap, 537, A16

\bibitem[{{Fontanot} et~al.(2014){Fontanot}, {Cristiani}, {Pfrommer}, {Cupani}
  \& {Vanzella}}]{Fontanot14}
{Fontanot} F., {Cristiani} S., {Pfrommer} C., {Cupani} G., {Vanzella} E., 2014,
  \mnras, 438, 2097

\bibitem[{{Fujita} et~al.(2003){Fujita}, {Martin}, {Mac Low} \&
  {Abel}}]{Fujita03}
{Fujita} A., {Martin} C.~L., {Mac Low} M.~M., {Abel} T., 2003, \apj, 599, 50

\bibitem[{{Furlanetto} \& {Dixon}(2010)}]{Furlanetto10}
{Furlanetto} S.~R., {Dixon} K.~L., 2010, \apj, 714, 355

\bibitem[{{Georgakakis} et~al.(2015)}]{Georgakakis15}
{Georgakakis} A. et~al., 2015, \mnras, 453, 1946

\bibitem[{{Giallongo} et~al.(2012){Giallongo}, {Menci}, {Fiore}, {Castellano},
  {Fontana}, {Grazian} \& {Pentericci}}]{Giallongo12}
{Giallongo} E., {Menci} N., {Fiore} F., {Castellano} M., {Fontana} A.,
  {Grazian} A., {Pentericci} L., 2012, \apj, 755, 124

\bibitem[{{Giallongo} et~al.(2015)}]{Giallongo15}
{Giallongo} E. et~al., 2015, \aap, 578, A83

\bibitem[{{Glikman} et~al.(2011){Glikman}, {Djorgovski}, {Stern}, {Dey},
  {Jannuzi} \& {Lee}}]{Glikman11}
{Glikman} E., {Djorgovski} S.~G., {Stern} D., {Dey} A., {Jannuzi} B.~T., {Lee}
  K.~S., 2011, \apjl, 728, L26

\bibitem[{{Gnedin} et~al.(2008){Gnedin}, {Kravtsov} \& {Chen}}]{Gnedin08}
{Gnedin} N.~Y., {Kravtsov} A.~V., {Chen} H.~W., 2008, \apj, 672, 765

\bibitem[{{Gordon} et~al.(2003){Gordon}, {Clayton}, {Misselt}, {Landolt} \&
  {Wolff}}]{Gordon03}
{Gordon} K.~D., {Clayton} G.~C., {Misselt} K.~A., {Landolt} A.~U., {Wolff}
  M.~J., 2003, \apj, 594, 279

\bibitem[{{Goto} et~al.(2011){Goto}, {Utsumi}, {Hattori}, {Miyazaki} \&
  {Yamauchi}}]{Goto11}
{Goto} T., {Utsumi} Y., {Hattori} T., {Miyazaki} S., {Yamauchi} C., 2011,
  \mnras, 415, L1

\bibitem[{{Grazian} et~al.(2016)}]{Grazian15}
{Grazian} A. et~al., 2016, \aap, 585, A48

\bibitem[{{Grimes} et~al.(2009)}]{Grimes09}
{Grimes} J.~P. et~al., 2009, \apjs, 181, 272

\bibitem[{{Haardt} \& {Madau}(1996)}]{HM96}
{Haardt} F., {Madau} P., 1996, \apj, 461, 20

\bibitem[{{Haardt} \& {Madau}(2012)}]{HM12}
{Haardt} F., {Madau} P., 2012, \apj, 746, 125

\bibitem[{{Haardt} \& {Salvaterra}(2015)}]{Haardt15}
{Haardt} F., {Salvaterra} R., 2015, \aap, 575, L16

\bibitem[{{Hinshaw} et~al.(2013)}]{Hinshaw13}
{Hinshaw} G. et~al., 2013, \apjs, 208, 19

\bibitem[{{Inoue} et~al.(2006){Inoue}, {Iwata} \& {Deharveng}}]{Inoue06}
{Inoue} A.~K., {Iwata} I., {Deharveng} J.~M., 2006, \mnras, 371, L1

\bibitem[{{Inoue} et~al.(2014){Inoue}, {Shimizu}, {Iwata} \&
  {Tanaka}}]{InoueAK14}
{Inoue} A.~K., {Shimizu} I., {Iwata} I., {Tanaka} M., 2014, \mnras, 442, 1805

\bibitem[{{Iwata} et~al.(2009)}]{Iwata09}
{Iwata} I. et~al., 2009, \apj, 692, 1287

\bibitem[{{Kashikawa} et~al.(2015)}]{Kashikawa15}
{Kashikawa} N. et~al., 2015, \apj, 798, 28

\bibitem[{{Khaire} \& {Srianand}(2013)}]{Khaire13}
{Khaire} V., {Srianand} R., 2013, \mnras, 431, L53

\bibitem[{{Khaire} \& {Srianand}(2015{\natexlab{a}})}]{Khaire15puc}
{Khaire} V., {Srianand} R., 2015{\natexlab{a}}, \mnras, 451, L30

\bibitem[{{Khaire} \& {Srianand}(2015{\natexlab{b}})}]{Khaire15ebl}
{Khaire} V., {Srianand} R., 2015{\natexlab{b}}, \apj, 805, 33

\bibitem[{{Kim} et~al.(2013){Kim}, {Partl}, {Carswell} \& {M{\"u}ller}}]{Kim13}
{Kim} T.~S., {Partl} A.~M., {Carswell} R.~F., {M{\"u}ller} V., 2013, \aap, 552,
  A77

\bibitem[{{Kimm} \& {Cen}(2014)}]{Kimm14}
{Kimm} T., {Cen} R., 2014, \apj, 788, 121

\bibitem[{{Kollmeier} et~al.(2014)}]{Kollmeier14}
{Kollmeier} J.~A. et~al., 2014, \apjl, 789, L32

\bibitem[{{Kroupa}(2001)}]{Kroupa01}
{Kroupa} P., 2001, \mnras, 322, 231

\bibitem[{{Kuhlen} \& {Faucher-Gigu{\`e}re}(2012)}]{Kuhlen12}
{Kuhlen} M., {Faucher-Gigu{\`e}re} C.~A., 2012, \mnras, 423, 862

\bibitem[{{Larson} et~al.(2011)}]{Larson11}
{Larson} D. et~al., 2011, \apjs, 192, 16

\bibitem[{{Leitet} et~al.(2013){Leitet}, {Bergvall}, {Hayes}, {Linn{\'e}} \&
  {Zackrisson}}]{Leitet13}
{Leitet} E., {Bergvall} N., {Hayes} M., {Linn{\'e}} S., {Zackrisson} E., 2013,
  \aap, 553, A106

\bibitem[{{Leitherer} et~al.(1999)}]{Leitherer99}
{Leitherer} C. et~al., 1999, \apjs, 123, 3

\bibitem[{{Lusso} et~al.(2015){Lusso}, {Worseck}, {Hennawi}, {Prochaska},
  {Vignali}, {Stern} \& {O'Meara}}]{Lusso15}
{Lusso} E., {Worseck} G., {Hennawi} J.~F., {Prochaska} J.~X., {Vignali} C.,
  {Stern} J., {O'Meara} J.~M., 2015, \mnras, 449, 4204

\bibitem[{{Lusso} et~al.(2010)}]{Lusso10}
{Lusso} E. et~al., 2010, \aap, 512, A34

\bibitem[{{Ma} et~al.(2015){Ma}, {Kasen}, {Hopkins}, {Faucher-Gigu{\`e}re},
  {Quataert}, {Kere{\v s}} \& {Murray}}]{Ma15}
{Ma} X., {Kasen} D., {Hopkins} P.~F., {Faucher-Gigu{\`e}re} C.~A., {Quataert}
  E., {Kere{\v s}} D., {Murray} N., 2015, \mnras, 453, 960

\bibitem[{{Madau} \& {Dickinson}(2014)}]{Madau14}
{Madau} P., {Dickinson} M., 2014, \araa, 52, 415

\bibitem[{{Madau} \& {Haardt}(2015)}]{Madau15}
{Madau} P., {Haardt} F., 2015, \apjl, 813, L8

\bibitem[{{Madau} et~al.(1999){Madau}, {Haardt} \& {Rees}}]{Madau99}
{Madau} P., {Haardt} F., {Rees} M.~J., 1999, \apj, 514, 648

\bibitem[{{Mashian} et~al.(2016){Mashian}, {Oesch} \& {Loeb}}]{Mashian15}
{Mashian} N., {Oesch} P.~A., {Loeb} A., 2016, \mnras, 455, 2101

\bibitem[{{Masters} et~al.(2012)}]{Masters12}
{Masters} D. et~al., 2012, \apj, 755, 169

\bibitem[{{McGreer} et~al.(2015){McGreer}, {Mesinger} \&
  {D'Odorico}}]{McGreer15}
{McGreer} I.~D., {Mesinger} A., {D'Odorico} V., 2015, \mnras, 447, 499

\bibitem[{{McGreer} et~al.(2013)}]{McGreer13}
{McGreer} I.~D. et~al., 2013, \apj, 768, 105

\bibitem[{{McLeod} et~al.(2015){McLeod}, {McLure}, {Dunlop}, {Robertson},
  {Ellis} \& {Targett}}]{McLeod15}
{McLeod} D.~J., {McLure} R.~J., {Dunlop} J.~S., {Robertson} B.~E., {Ellis}
  R.~S., {Targett} T.~A., 2015, \mnras, 450, 3032

\bibitem[{{Meiksin} \& {White}(2003)}]{Meiksin03}
{Meiksin} A., {White} M., 2003, \mnras, 342, 1205

\bibitem[{{Merloni} et~al.(2014)}]{Merloni14}
{Merloni} A. et~al., 2014, \mnras, 437, 3550

\bibitem[{{Mesinger} \& {Furlanetto}(2008)}]{Mesinger08}
{Mesinger} A., {Furlanetto} S.~R., 2008, \mnras, 386, 1990

\bibitem[{{Micheva} et~al.(2015){Micheva}, {Iwata}, {Inoue}, {Matsuda},
  {Yamada} \& {Hayashino}}]{Micheva15}
{Micheva} G., {Iwata} I., {Inoue} A.~K., {Matsuda} Y., {Yamada} T., {Hayashino}
  T., 2015, ArXiv e-prints: 1509.03996

\bibitem[{{Miralda-Escude} \& {Ostriker}(1990)}]{Miralda90}
{Miralda-Escude} J., {Ostriker} J.~P., 1990, \apj, 350, 1

\bibitem[{{Mitra} et~al.(2013){Mitra}, {Ferrara} \& {Choudhury}}]{Mitra13}
{Mitra} S., {Ferrara} A., {Choudhury} T.~R., 2013, \mnras, 428, L1

\bibitem[{{Mitra} et~al.(2015){Mitra}, {Choudhury} \& {Ferrara}}]{Mitra15}
{Mitra} S., {Choudhury} T.~R., {Ferrara} A., 2015, \mnras, 454, L76

\bibitem[{{Moretti} et~al.(2012)}]{Moretti12}
{Moretti} A., {Vattakunnel} S., {Tozzi} P., {Salvaterra} R., {Severgnini} P.,
  {Fugazza} D., {Haardt} F., {Gilli} R., 2012, \aap, 548, A87

\bibitem[{{Mostardi} et~al.(2015){Mostardi}, {Shapley}, {Steidel}, {Trainor},
  {Reddy} \& {Siana}}]{Mostardi15}
{Mostardi} R.~E., {Shapley} A.~E., {Steidel} C.~C., {Trainor} R.~F., {Reddy}
  N.~A., {Siana} B., 2015, \apj, 810, 107

\bibitem[{{Nandra} \& {Pounds}(1994)}]{Nandra94}
{Nandra} K., {Pounds} K.~A., 1994, \mnras, 268, 405

\bibitem[{{Nestor} et~al.(2013){Nestor}, {Shapley}, {Kornei}, {Steidel} \&
  {Siana}}]{Nestor13}
{Nestor} D.~B., {Shapley} A.~E., {Kornei} K.~A., {Steidel} C.~C., {Siana} B.,
  2013, \apj, 765, 47

\bibitem[{{Noterdaeme} et~al.(2009){Noterdaeme}, {Ledoux}, {Srianand},
  {Petitjean} \& {Lopez}}]{Noterdaeme09}
{Noterdaeme} P., {Ledoux} C., {Srianand} R., {Petitjean} P., {Lopez} S., 2009,
  \aap, 503, 765

\bibitem[{{Noterdaeme} et~al.(2012)}]{Noterdaeme12}
{Noterdaeme} P. et~al., 2012, \aap, 547, L1

\bibitem[{{Oesch} et~al.(2014)}]{Oesch14}
{Oesch} P.~A. et~al., 2014, \apj, 786, 108

\bibitem[{{O'Meara} et~al.(2007){O'Meara}, {Prochaska}, {Burles}, {Prochter},
  {Bernstein} \& {Burgess}}]{OMeara07}
{O'Meara} J.~M., {Prochaska} J.~X., {Burles} S., {Prochter} G., {Bernstein}
  R.~A., {Burgess} K.~M., 2007, \apj, 656, 666

\bibitem[{{O'Meara} et~al.(2013){O'Meara}, {Prochaska}, {Worseck}, {Chen} \&
  {Madau}}]{OMeara13}
{O'Meara} J.~M., {Prochaska} J.~X., {Worseck} G., {Chen} H.~W., {Madau} P.,
  2013, \apj, 765, 137

\bibitem[{{Paardekooper} et~al.(2011){Paardekooper}, {Pelupessy}, {Altay} \&
  {Kruip}}]{Paardekooper11}
{Paardekooper} J.~P., {Pelupessy} F.~I., {Altay} G., {Kruip} C.~J.~H., 2011,
  \aap, 530, A87

\bibitem[{{Padmanabhan} et~al.(2014){Padmanabhan}, {Choudhury} \&
  {Srianand}}]{Padmanabhan14}
{Padmanabhan} H., {Choudhury} T.~R., {Srianand} R., 2014, \mnras, 443, 3761

\bibitem[{{Padmanabhan}(2002)}]{Paddy3}
{Padmanabhan} T., 2002, {Theoretical Astrophysics - Volume 3, Galaxies and
  Cosmology}

\bibitem[{{Palanque-Delabrouille} et~al.(2013)}]{Palanque13}
{Palanque-Delabrouille} N. et~al., 2013, \aap, 551, A29

\bibitem[{{Paresce} et~al.(1980){Paresce}, {McKee} \& {Bowyer}}]{Paresce}
{Paresce} F., {McKee} C.~F., {Bowyer} S., 1980, \apj, 240, 387

\bibitem[{{Peebles}(1993)}]{Peebles93}
{Peebles} P.~J.~E., 1993, {Principles of Physical Cosmology}

\bibitem[{{Planck Collaboration} et~al.(2015)}]{Planck15}
{Planck Collaboration} et~al., 2015, arXiv:1502.01589

\bibitem[{{Prochaska} et~al.(2014){Prochaska}, {Madau}, {O'Meara} \&
  {Fumagalli}}]{Prochaska14}
{Prochaska} J.~X., {Madau} P., {O'Meara} J.~M., {Fumagalli} M., 2014, \mnras,
  438, 476

\bibitem[{{Rahmati} et~al.(2013){Rahmati}, {Pawlik}, {Raicevic} \&
  {Schaye}}]{Rahmati13}
{Rahmati} A., {Pawlik} A.~H., {Raicevic} M., {Schaye} J., 2013, \mnras, 430,
  2427

\bibitem[{{Rhoads} et~al.(2004)}]{Rhoads04}
{Rhoads} J.~E. et~al., 2004, \apj, 611, 59

\bibitem[{{Ricotti} \& {Shull}(2000)}]{Ricotti00}
{Ricotti} M., {Shull} J.~M., 2000, \apj, 542, 548

\bibitem[{{Robertson} et~al.(2015){Robertson}, {Ellis}, {Furlanetto} \&
  {Dunlop}}]{Robertson15}
{Robertson} B.~E., {Ellis} R.~S., {Furlanetto} S.~R., {Dunlop} J.~S., 2015,
  \apjl, 802, L19

\bibitem[{{Ross} et~al.(2013)}]{Ross13}
{Ross} N.~P. et~al., 2013, \apj, 773, 14

\bibitem[{{Roy} et~al.(2015){Roy}, {Nath} \& {Sharma}}]{Roy15}
{Roy} A., {Nath} B.~B., {Sharma} P., 2015, \mnras, 451, 1939

\bibitem[{{Salpeter}(1955)}]{Salpeter55}
{Salpeter} E.~E., 1955, \apj, 121, 161

\bibitem[{{Samui} et~al.(2007){Samui}, {Srianand} \& {Subramanian}}]{Samui07}
{Samui} S., {Srianand} R., {Subramanian} K., 2007, \mnras, 377, 285

\bibitem[{{Schaye}(2001)}]{Schaye01}
{Schaye} J., 2001, \apj, 559, 507

\bibitem[{{Schenker} et~al.(2014){Schenker}, {Ellis}, {Konidaris} \&
  {Stark}}]{Schenker14}
{Schenker} M.~A., {Ellis} R.~S., {Konidaris} N.~P., {Stark} D.~P., 2014, \apj,
  795, 20

\bibitem[{{Schulze} et~al.(2009){Schulze}, {Wisotzki} \&
  {Husemann}}]{Schulze09}
{Schulze} A., {Wisotzki} L., {Husemann} B., 2009, \aap, 507, 781

\bibitem[{{Shapiro} et~al.(1994){Shapiro}, {Giroux} \& {Babul}}]{Shapiro94}
{Shapiro} P.~R., {Giroux} M.~L., {Babul} A., 1994, \apj, 427, 25

\bibitem[{{Shapley} et~al.(2006){Shapley}, {Steidel}, {Pettini}, {Adelberger}
  \& {Erb}}]{Shapley06}
{Shapley} A.~E., {Steidel} C.~C., {Pettini} M., {Adelberger} K.~L., {Erb}
  D.~K., 2006, \apj, 651, 688

\bibitem[{{Shull} et~al.(1999){Shull}, {Roberts}, {Giroux}, {Penton} \&
  {Fardal}}]{Shull99}
{Shull} J.~M., {Roberts} D., {Giroux} M.~L., {Penton} S.~V., {Fardal} M.~A.,
  1999, \aj, 118, 1450

\bibitem[{{Shull} et~al.(2012){Shull}, {Harness}, {Trenti} \&
  {Smith}}]{Shull12}
{Shull} J.~M., {Harness} A., {Trenti} M., {Smith} B.~D., 2012, \apj, 747, 100

\bibitem[{{Shull} et~al.(2015){Shull}, {Moloney}, {Danforth} \&
  {Tilton}}]{Shull15}
{Shull} J.~M., {Moloney} J., {Danforth} C.~W., {Tilton} E.~M., 2015, \apj, 811,
  3

\bibitem[{{Siana} et~al.(2010)}]{Siana10}
{Siana} B. et~al., 2010, \apj, 723, 241

\bibitem[{{Siana} et~al.(2015)}]{Siana15}
{Siana} B. et~al., 2015, \apj, 804, 17

\bibitem[{{Springel}(2005)}]{Springel05}
{Springel} V., 2005, \mnras, 364, 1105

\bibitem[{{Stalin} et~al.(2010){Stalin}, {Petitjean}, {Srianand}, {Fox},
  {Coppolani} \& {Schwope}}]{Stalin10}
{Stalin} C.~S., {Petitjean} P., {Srianand} R., {Fox} A.~J., {Coppolani} F.,
  {Schwope} A., 2010, \mnras, 401, 294

\bibitem[{{Steidel} et~al.(2001){Steidel}, {Pettini} \&
  {Adelberger}}]{Steidel01}
{Steidel} C.~C., {Pettini} M., {Adelberger} K.~L., 2001, \apj, 546, 665

\bibitem[{{Stevans} et~al.(2014){Stevans}, {Shull}, {Danforth} \&
  {Tilton}}]{Stevans14}
{Stevans} M.~L., {Shull} J.~M., {Danforth} C.~W., {Tilton} E.~M., 2014, \apj,
  794, 75

\bibitem[{{Topping} \& {Shull}(2015)}]{Topping15}
{Topping} M.~W., {Shull} J.~M., 2015, \apj, 800, 97

\bibitem[{{Tumlinson} et~al.(2001){Tumlinson}, {Giroux} \&
  {Shull}}]{Tumlinson01}
{Tumlinson} J., {Giroux} M.~L., {Shull} J.~M., 2001, \apjl, 550, L1

\bibitem[{{Vangioni} et~al.(2015){Vangioni}, {Olive}, {Prestegard}, {Silk},
  {Petitjean} \& {Mandic}}]{Vangioni15}
{Vangioni} E., {Olive} K.~A., {Prestegard} T., {Silk} J., {Petitjean} P.,
  {Mandic} V., 2015, \mnras, 447, 2575

\bibitem[{{Vanzella} et~al.(2010)}]{Vanzella10esc}
{Vanzella} E. et~al., 2010, \apj, 725, 1011

\bibitem[{{Venkatesan} et~al.(2001){Venkatesan}, {Giroux} \&
  {Shull}}]{Venkatesan01}
{Venkatesan} A., {Giroux} M.~L., {Shull} J.~M., 2001, \apj, 563, 1

\bibitem[{{Wakker} et~al.(2015){Wakker}, {Hernandez}, {French}, {Kim},
  {Oppenheimer} \& {Savage}}]{Wakker15}
{Wakker} B.~P., {Hernandez} A.~K., {French} D.~M., {Kim} T.~S., {Oppenheimer}
  B.~D., {Savage} B.~D., 2015, \apj, 814, 40

\bibitem[{{Weigel} et~al.(2015){Weigel}, {Schawinski}, {Treister}, {Urry},
  {Koss} \& {Trakhtenbrot}}]{Weigel15}
{Weigel} A.~K., {Schawinski} K., {Treister} E., {Urry} C.~M., {Koss} M.,
  {Trakhtenbrot} B., 2015, \mnras, 448, 3167

\bibitem[{{Worseck} et~al.(2014{\natexlab{a}}){Worseck}, {Prochaska}, {Hennawi}
  \& {McQuinn}}]{Worseck14}
{Worseck} G., {Prochaska} J.~X., {Hennawi} J.~F., {McQuinn} M.,
  2014{\natexlab{a}}, ArXiv e-prints: 1405.7405

\bibitem[{{Worseck} et~al.(2014{\natexlab{b}})}]{Worseck14LLS}
{Worseck} G. et~al., 2014{\natexlab{b}}, \mnras, 445, 1745

\bibitem[{{Wyithe} \& {Bolton}(2011)}]{Wyithe11}
{Wyithe} J.~S.~B., {Bolton} J.~S., 2011, \mnras, 412, 1926

\bibitem[{{Yajima} et~al.(2009){Yajima}, {Umemura}, {Mori} \&
  {Nakamoto}}]{Yajima09}
{Yajima} H., {Umemura} M., {Mori} M., {Nakamoto} T., 2009, \mnras, 398, 715

\end{thebibliography}

\appendix
\section{Photoionization rates and clumping factor}\label{app}
%
\begin{figure*}
\centering
\includegraphics[totalheight=0.7\textheight, trim=7.5cm 0cm 0cm 0cm, clip=true, angle=90]{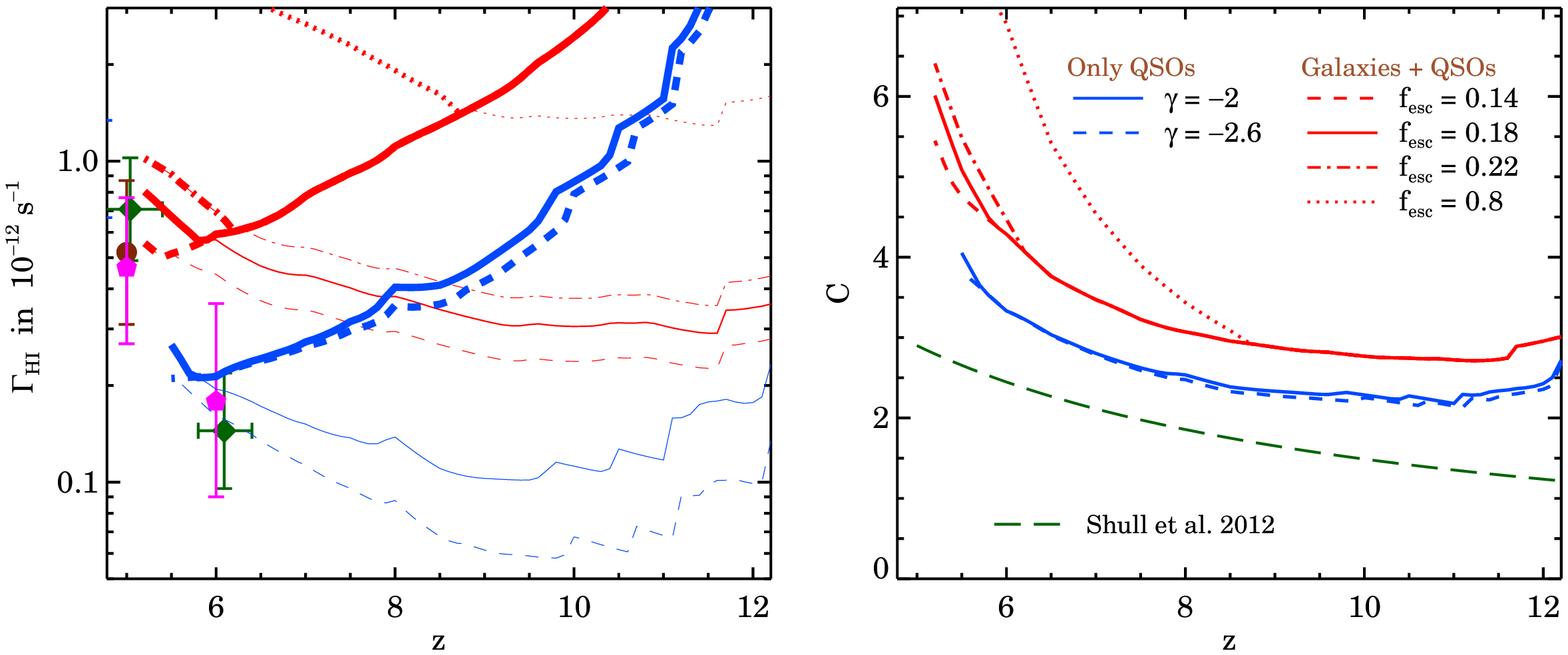}
\caption{\emph{Left}: The mean values of $\Gamma_{\rm HI}(z)$ within the H~{\sc ii} 
bubbles (\emph{thick curves}) are shown for different models of reionization discussed in 
Section~\ref{sec4.2} (Galaxies + QSO models) and Section~\ref{sec4.4} (models with only QSOs).
Legends in the \emph{right-hand panel} describe various models. The measurements of 
$\Gamma_{\rm HI}(z)$ are the same as shown in the right-hand panel of Fig.~\ref{fig3}. 
Thin curves represent $Q_{\rm HII}(z)\times \Gamma_{\rm HI}(z)$ which 
is equivalent to the definition of photoionization rate by \citetalias{HM12}.
\emph{Right}: The $C(z)$ is shown corresponding to the $\Gamma_{\rm HI}(z)$ within the H~{\sc ii} 
bubbles (as shown in the \emph{left-hand panel}). For comparison, we also show the mean 
value of $C(z)$ obtained by \citet{Shull12}.}
\label{fig.A1}
\end{figure*}
%
 In Fig.~\ref{fig.A1}, we show $\Gamma_{\rm HI}(z)$ 
(\emph{thick curves in left-hand panel}) and
$C(z)$ (\emph{right-hand panel}) within the H~{\sc ii} bubbles 
for the models of reionization presented in Section~\ref{sec4.2} and \ref{sec4.4}. 
For the models of reionization driven by galaxies with constant $f_{\rm esc}(z>z_{\rm re})$ 
(Section~\ref{sec4.2}), the $\Gamma_{\rm HI}(z)$ is independent of the $f_{\rm esc}$ values
in the pre-reionization era. This is an artefact of using constant $f_{\rm esc}(z>z_{\rm re})$. 
Under this assumption, the $Q_{\rm HII}$ at any $z$ in pre-reionization era scales 
with the value of $f_{\rm esc}$, as evident from Eq.~(\ref{Eq.q}). 
In the same way, the $J_{\nu}$ scales with the $f_{\rm esc}/Q_{\rm HII}(z)$. 
Therefore, the $\Gamma_{\rm HI}(z)$ becomes independent of the $f_{\rm esc}$
in the pre-reionization era. 

The $\Gamma_{\rm HI}(z)$ values for models of reionization driven by galaxies are consistent with the
measurements at $z\sim5$ but off by 1 to 2-$\sigma$ level at $z\sim6$. 
$\Gamma_{\rm HI}(z)$ for the models of reionization by 
QSOs alone (Section~\ref{sec4.4}) are consistent with the measurements at $z\sim6$.
Main reason for this difference in $\Gamma_{\rm HI}(z)$ in the galaxy dominated 
and QSO dominated models of reionization is arising
because of the different SEDs of the galaxies and QSOs. 
The SED for QSOs (with $\alpha=-1.4$) is more flat than the SED of galaxies ($\beta=-1.8$). 
Therefore, to obtain the same $\dot n$, galaxies need more photons at H~{\sc i} ionizing
edge as compared to QSOs. This eventually leads to a higher $\Gamma_{\rm HI}$, due to the $\nu^{-3}$ 
dependence of $\sigma_{\rm HI}(\nu)$, in the models where galaxies dominate the reionization.
In the \emph{left-hand panel} of Fig.~\ref{fig.A1} we also show the $Q_{\rm HII}(z)\times \Gamma_{\rm HI}(z)$ 
(\emph{thin curves}), that is equivalent to the definition of photoionization 
rate by \citetalias{HM12}.

The corresponding $C(z)$, shown in the \emph{right-hand panel} of Fig.~\ref{fig.A1}, 
is calculated in the simulation box at different $z$. 
It is then extrapolated at $z$ where we do not have the stored simulation boxes. 
The $C(z)$ depends on the $\Gamma_{\rm HI}(z)$ within the H~{\sc ii} bubbles. 
Therefore, for galaxy dominated reionization models, $C(z)$ is independent
of the values of $f_{\rm esc}$ in the pre-reionization era.
For comparison, we also show the mean value of $C(z)$ obtained by \citet{Shull12} 
which is systematically lower that what we obtain for both cases.

\end{document}